\documentclass[]{aa} 
 
\usepackage{natbib}  

\usepackage{graphicx}
\usepackage[varg]{txfonts}
\usepackage{hyperref}

\def\Teff  {$T_\mathrm{eff}$}
\def\logg  {$\log g$}

\def\vt    {$\rm v_{t}$}
\def\kms   {$\rm km\,s^{-1}$}

\begin{document}

\title{
Abundance patterns of the light neutron-capture elements in very and extremely metal-poor stars
\thanks{Based on observations collected at the European Organisation for Astronomical Research in the Southern Hemisphere under ESO programme  165.N-0276(A), (PI R.Cayrel). }
%\fnmsep
%\thanks{toto}
}
\author {
F. Spite\inst{1}\and 
M. Spite\inst{1}\and
B. Barbuy\inst{2}\and
P. Bonifacio\inst{1}\and
E. Caffau\inst{1}\and
P. Fran\c cois\inst{1,3}
%XXX\inst{1}\and 
%YYY\inst{2,1}\and
%ZZZ\inst{3}\and
 }

\institute {
GEPI, Observatoire de Paris, PSL Research University, CNRS,
Place Jules Janssen, 92190 Meudon, France
\and
Universidade de S\~ao Paulo, IAG, Rua do Mat\~ao 1226, Cidade Universit\'aria, 05508-900, S\~ao Paulo, Brazil
\and
Universit\'e de Picardie Jules Verne, 33 rue St-Leu, 80080, Amiens, France
%GEPI, Observatoire de Paris, PSL Research University, CNRS,
%Place Jules Janssen, 92190 Meudon, France
%\and
%tralala 
%\and
%Universidade de S\~ao Paulo, IAG, Rua do Mat\~ao 1226, Cidade Universit\'aria, 05508-900, S\~ao Paulo, Brazil
}

%   \date{Received September 15, 1996; accepted March 16, 1997}

\authorrunning{Spite et al.}

\titlerunning{Light neutron-capture elements in extremely metal-poor stars}

  \abstract
  % context heading (optional), leave it empty if necessary
{}
{The abundance patterns of the neutron-capture elements in metal-poor stars provide a unique record of the nucleosynthesis products of the earlier massive primitive objects.  
}
{We measured new abundances of so-called light neutron-capture of first peak elements using local thermodynamic equilibrium (LTE) 1D analysis; this  analysis resulted in a sample of 11 very metal-poor stars, from [Fe/H]=--2.5 to [Fe/H]=--3.4, and one carbon-rich star, CS\,22949-037 with [Fe/H]=--4.0. The abundances were compared to those observed in two classical metal-poor stars: the typical r-rich star CS\,31082-001 ([Eu/Fe] > +1.0) and the r-poor star HD\,122563 ([Eu/Fe] < 0.0), which are known to present a strong enrichment of the first peak neutron-capture elements relative to the second peak.
}
{Within the first peak, the abundances are well correlated in analogy to the well-known correlation inside the abundances of the second-peak elements. 
%\LEt{Please avoid the use of quotation marks when you are not actually quoting something. Please also avoid the use of italics used for emphasis. See Sect. 1.2 of the Author's Guide. Please check for these throughout and correct accordingly. (http://www.aanda.org/author-information/language-editing/1-introduction.}
In contrast, there is no correlation between any first peak element with any second peak element. We show that the scatter of the ratio of the first peak abundance over second peak abundance increases when the mean abundance of the second peak elements decreases from r-rich to r-poor stars.\\ 
We found two new r-poor stars that are very similar to HD\,122563. A third r-poor star, CS\,22897-008, is even more extreme; this star shows the most extreme example of first peak elements enrichment to date.  On the contrary, another r-poor star (BD--18\,5550) has a pattern of first peak elements that is similar to the typical r-rich stars CS\,31082-001,  however this star has some Mo enrichment.  
}
{The distribution of the neutron-capture elements in our very metal-poor stars can be understood as the combination of at least two mechanisms: one that enriches the forming stars cloud  homogeneously through the main r-process and leads to an element pattern similar to the r-rich stars, such as CS\,31082-001; and another that forms mainly lighter, first peak elements.
}


\maketitle

%
%---------------------------- Introduction -----------------------
\section{Introduction}
The neutron-capture elements are mainly synthesised, as recorded by the distribution of their isotopes in the Sun,  by two classical processes: either the slow {\sl(s)} or  rapid {\sl (r)} process.  Schematically, the heavy elements of the s-process could be built by slow neutron production of asymptotic giant branch stars (AGB) and the r-process elements could be built by a flux of high neutron density. 

The chemical composition of the old metal-poor stars reflects the production of elements in the first phases of galactic evolution. 
%\LEt{If you are referring to our local Milky Way Galaxy, use upper case for Galactic and Galaxy throughout. Please correct if necessary.}
The matter that formed these very old stars had been only enriched  by still older short-lived massive objects. However the exact nature of the sites of production of the observed neutron-capture elements in these stars is still actively debated \citep{CowanRS11}. The focus of neutron-capture abundance studies in the old metal-poor stars is an attempt to constrain the sites of the nucleosynthesis of these elements.

The r-process elements are produced by a rapid addition of neutrons to the atomic nuclei. This is the case in rather rare but productive  processes such as the mergers of neutron stars  \citep{KasenMB17,VangioniGD16,goriely15,wanajo14,FreiburghausRT99}  and of course in several kinds of explosive burning phases of supernovae (see e.g. \citet{WanajoJM11}, \citet{WintelerKP2012},  \citet{AokiIA17}, \citet{NishimuraST17},  \citet{BanerjeeQH17}). 

The s-process elements can be produced by the slow irradiations of abundant seed nuclei (e.g. iron nuclei) by the neutron flux delivered in either of two cases:\\

(i) \ By pulsating stars in a late evolutionary phase, i.e.  AGB and super AGB stars (SAGB) \citep{kappeler11,bisterzo17}. The initial mass of these {\bf SAGB} stars has to be lower than 9 $M_{\odot}$ to reach the AGB phase, but higher than about 6$M_{\odot}$ to have a lifetime short enough to be able to enrich the matter before the formation of the very old low mass stars observed today. 

(ii) In fast rotating massive stars (hereafter FRMS); see e.g. \citet{MeynetEM06}, \citet{HirschiMM07}.\\  

The results of the r- or s-processes are very different, and not only by the resulting isotopic composition.
For example, the neutrons slowly absorbed by seed nuclei in the FRMS process produce         a characteristic pattern with dominating lower mass elements in the first peak  \citep{FrischknechtHP16}; an example is given by \citet{YongND17}, who have analysed a star in the globular cluster $\Omega$ Cen.
  
In the old metal-poor Galactic stars, at a given metallicity, the scatter of the abundance of the neutron capture elements relative to iron [X/Fe]\footnote{We adopt the classical notation that, for each element X, $\rm A(X) = log(N_{X} / N_{H}) + 12$; $\rm [X/H]=  log(N_{X} / N_{H})_{star}-log(N_{X} / N_{H})_{Sun}$, and [X/Fe]=[X/H]--[Fe/H].}, is huge (compared to the low scatter of other  element ratios such as [Ca/Fe]), as noted for example by \citet{FrancoisDH07, SpiteS14}. 
Some stars are very rich in neutron-capture elements; these are called r-rich stars if [Eu/Fe]>0.3, but others, at the same metallicity, are r-poor with $\rm[Eu/Fe]\ll 0.0$.
It has been shown that the pattern of the element abundances in the r-rich stars is characteristic of the classical  r-process, \citep[see for example][]{CowanRS11}. But the pattern of the elements of the second peak ($\rm 56 \le Z \le 70$), is rather similar in r-rich and r-poor stars  \citep[see also][]{CowanRS11}.\\
 
On the contrary, the abundance pattern of the light neutron-capture elements (32 < Z < 56, or first peak elements) is very different in the r-rich and the r-poor stars. Compared to the classical r-rich metal-poor stars (for example CS\,31082-001 \citep{HillPC02,SiqueiraSB13} or CS\,22892-052 \citep{SnedenCL03},
some very metal-poor (VMP) stars such as HD\,122563 \citep{HondaAI06,HondaAI07,MontesBC07} present a strong excess of light neutron-capture elements \citep[see also: ][]{FrancoisDH07, Peterson11, Peterson13}. 
%\LEt{Please ensure that all abbreviations are written out at first mention, followed by the abbreviation in parentheses (even if you have already introduced them in the Abstract). After that please use only the abbreviation. Please check for your use of individual abbreviations throughout the paper. See Sect. 5.2.4 of Author's Guide (http://www.aanda.org/author-information/language-editing/1-introduction).}

The behaviour of the second peak elements and light neutron-capture elements Sr, Y, Zr (first peak elements with Z=38, 39, and 40) were presented for a homogeneous sample of old, very, and extremely metal-poor (EMP) stars  by \citet{FrancoisDH07}.
These stars were studied in the context of the ESO Large Programme ``First Stars'' (hereafter LP First Stars);  see \citet{CayrelDS04} and \citet{BonifacioSC09}.
The aim of the present paper is an attempt to determine in the same stars and the behaviour of the intermediate first peak elements from Mo to Ag (Z=42 to 47),  when possible.

\section {Observational data} 
The spectral lines of the light neutron-capture elements measured in this work are all located in the near UV between 338 and 390 nm.
The spectra were obtained with the Very Large Telescope (VLT)  and the spectrograph UVES \citep{DekkerDK00} as part of the LP First Stars. The resolving power $R$ was close to 47000 and the spectra were reduced using the UVES context \citep{BallesterMB00}; more information can be found in \citet{CayrelDS04}. 

%TABLE 1
\begin{table}
\begin{center}   
\caption[]{
For each star column 2 the S/N of the mean spectrum measured at 385 nm and in column 3 to 6 the parameters of the adopted models are listed.}
\label{sn}
\begin{tabular}{lccccrr}
\hline
  Object       &    S/N(385nm) &\Teff & \logg & \vt  & [M/H]\\ 
\\
HD\,2796       &        280    &4950  & 1.5   & 2.1  & -2.5 \\
HD\,186478     &        190    &4700  & 1.3   & 2.0  & -2.6 \\
BD--18\,5550   &        210    &4750  & 1.4   & 1.8  & -3.0 \\
BS\,17569-049  &        125    &4700  & 1.2   & 1.9  & -2.9 \\
CS\,22873-055  &        135    &4550  & 0.7   & 2.2  & -3.0 \\
CS\,22873-166  &        130    &4550  & 0.9   & 2.1  & -3.0 \\
CS\,22896-154  &        125    &5250  & 2.7   & 1.2  & -2.7 \\
CS\,22897-008  &        155    &4900  & 1.7   & 2.0  & -3.4 \\
CS\,22949-037  &        135    &4900  & 1.5   & 1.8  & -4.0 \\
CS\,22953-003  &        150    &5100  & 2.3   & 1.7  & -2.8 \\
CS\,22966-057  &        100    &5300  & 2.2   & 1.4  & -2.6 \\
CS\,29518-051  &        190    &5200  & 2.6   & 1.4  & -2.7 \\
\hline    &&&
\end{tabular}  
\end{center}   
\end{table}

% Figure 1
\begin{figure}
\resizebox{7cm}{3.3cm}                   
{\includegraphics {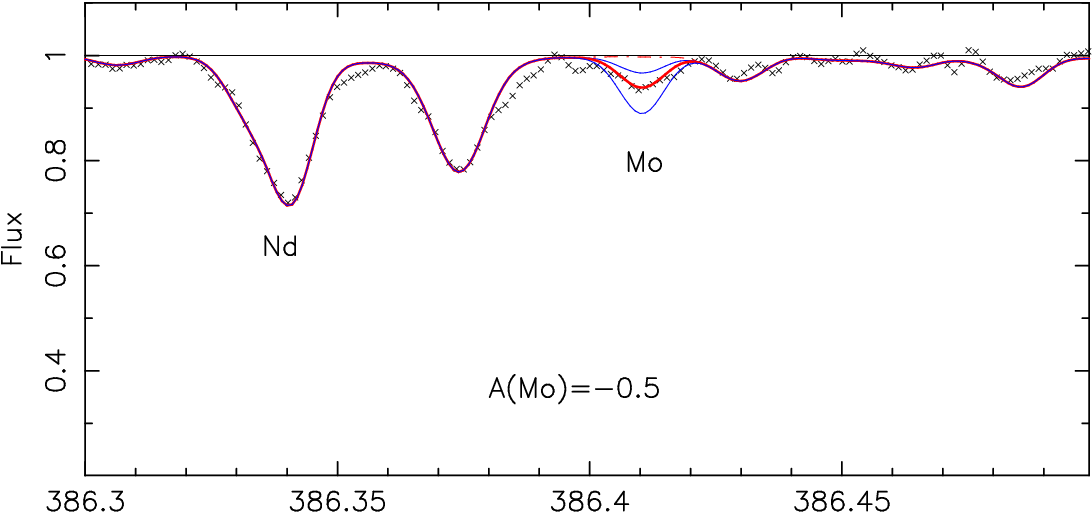}}
\resizebox{7cm}{3.3cm}                   
{\includegraphics {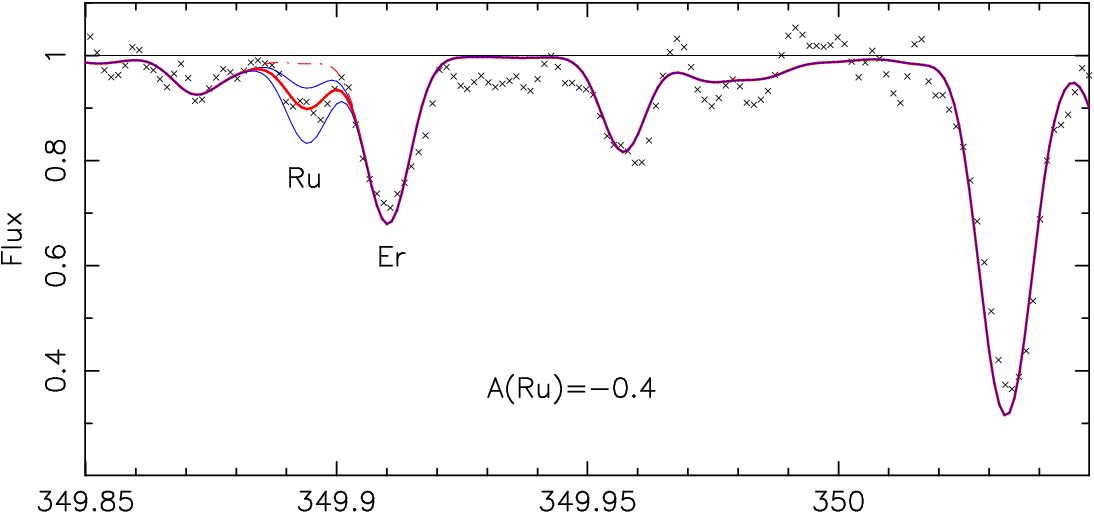}}
\resizebox{7cm}{3.3cm}                   
{\includegraphics {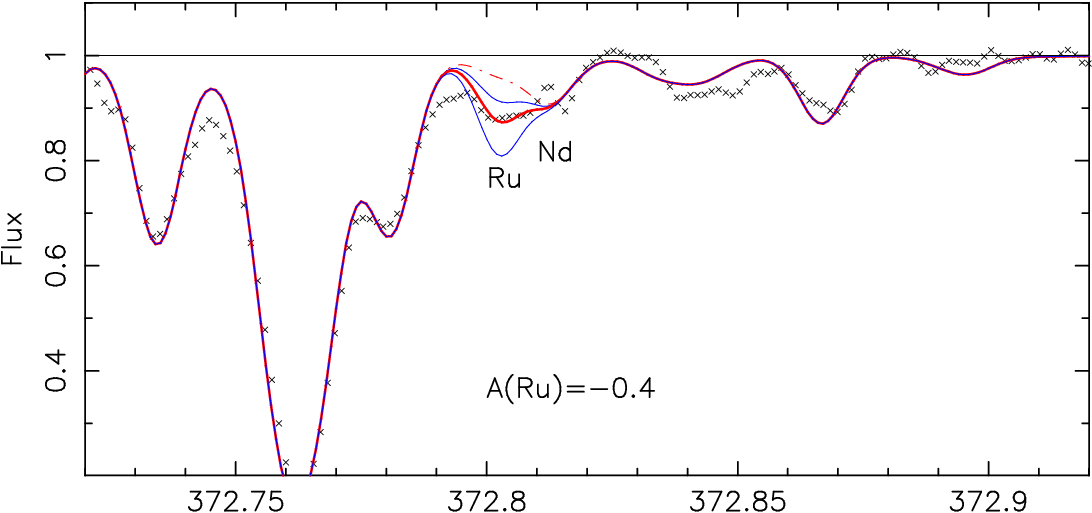}}
\resizebox{7cm}{3.3cm}                   
{\includegraphics {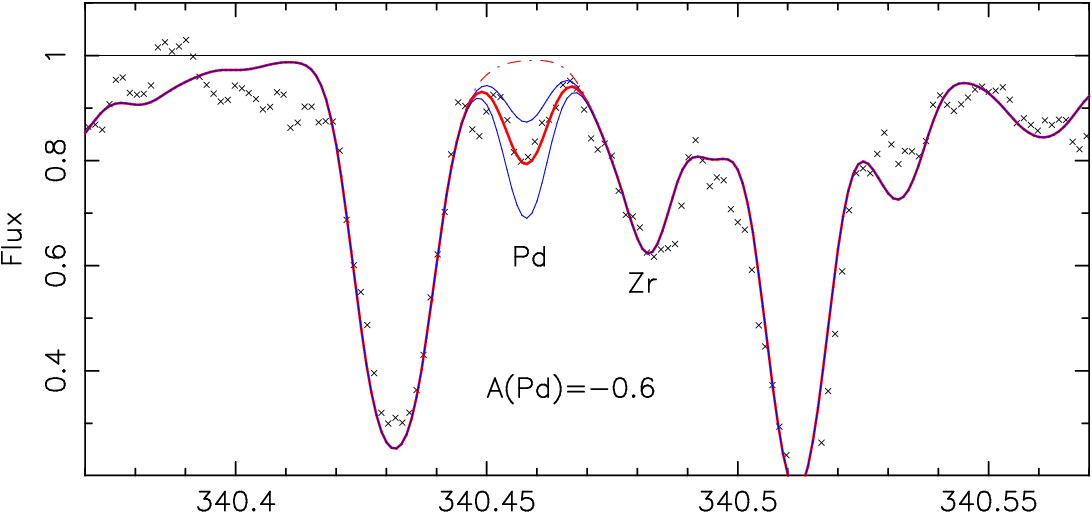}}
\resizebox{7cm}{3.3cm}                   
{\includegraphics {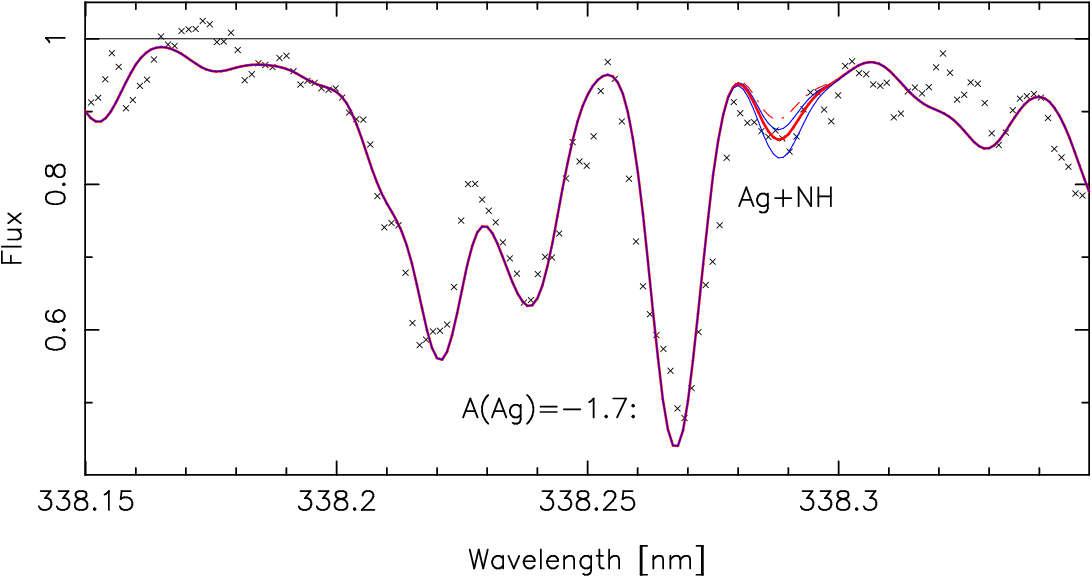}}
\caption[]{Observed profile of the lines of Mo, Ru, Pd, and Ag in BS~17569-049. Crosses indicate the observations. The synthetic profiles were computed with three options: an absence of the neutron-capture element (red dash-dotted line), the adopted abundance indicated in the figure (red thick line), and changing this adopted abundance by $\pm 0.3$ dex (blue thin lines). BS~17569-049 is a mixed star (therefore with a high abundance of nitrogen) and, as a consequence, the abundance of Ag deduced from the blended feature at 338.28\,nm is very uncertain.}
\label {fits}
\end{figure}

% Figure 2
\begin{figure*}
\begin{center}
\resizebox{5.6cm}{5.6cm}
{\includegraphics {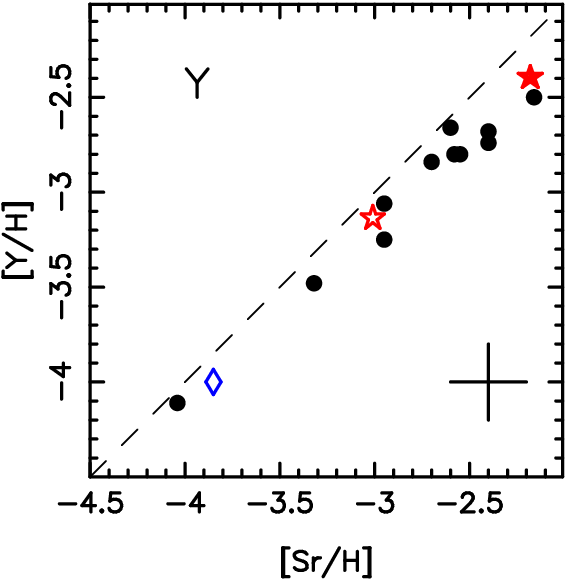}}
\resizebox{5.6cm}{5.6cm}  
{\includegraphics {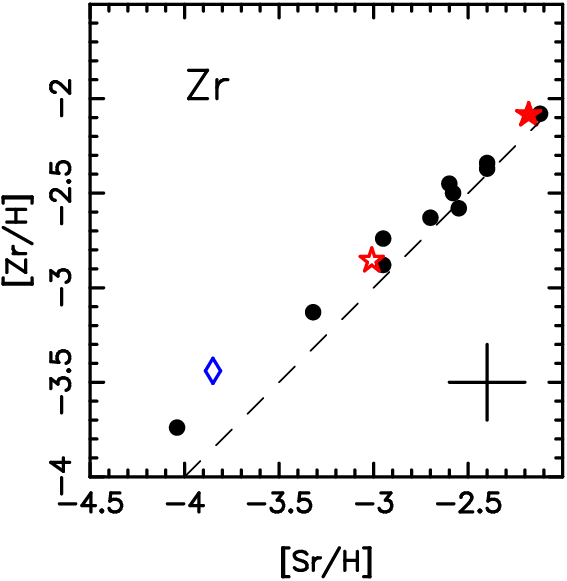}}
\resizebox{5.6cm}{5.6cm}  
{\includegraphics {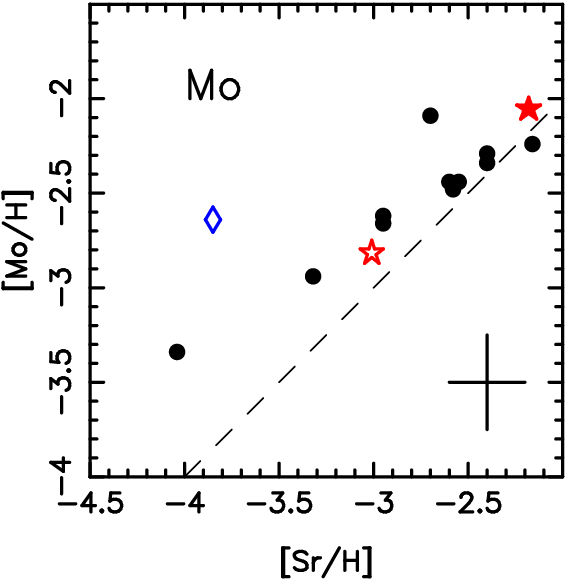}}
\resizebox{5.6cm}{5.6cm}  
{\includegraphics {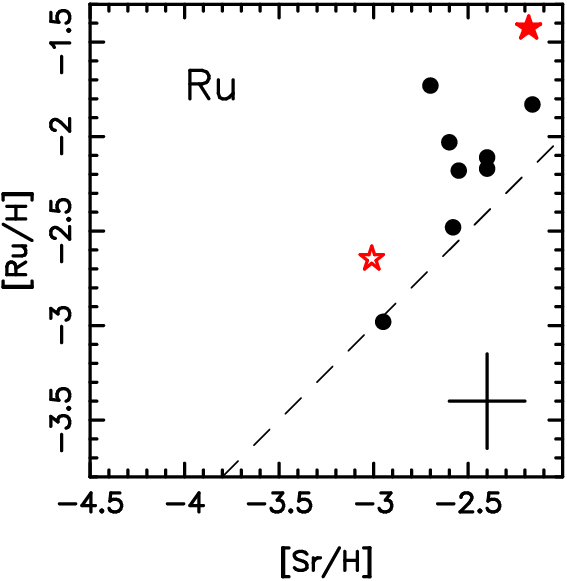}}
\resizebox{5.6cm}{5.6cm}  
{\includegraphics {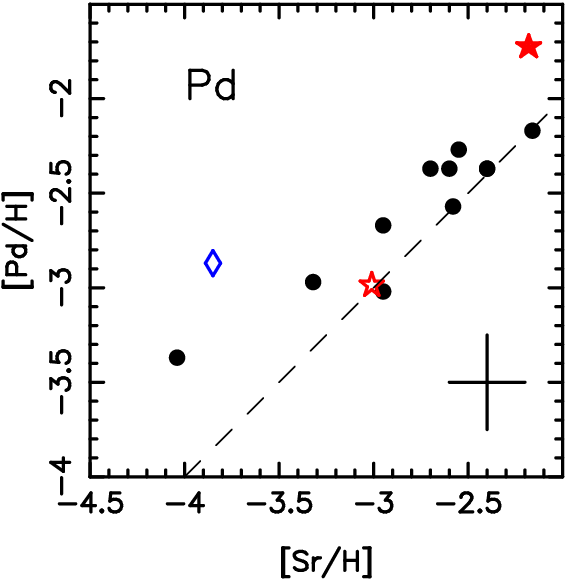}}
\resizebox{5.6cm}{5.6cm}  
{\includegraphics {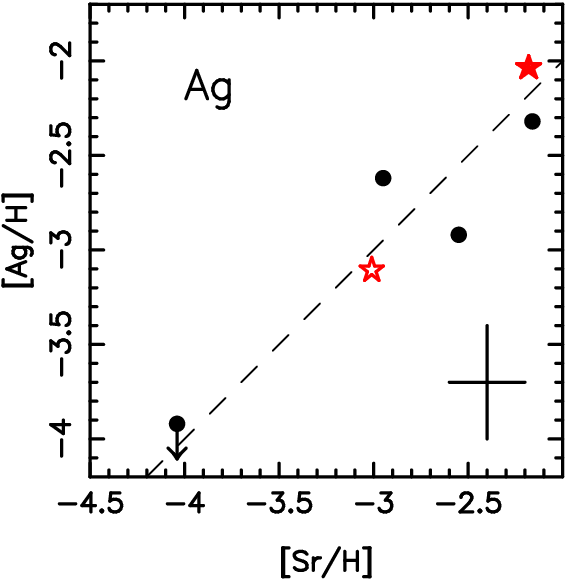}}
\caption[]{Six relations between the ratios [X/H] versus [Sr/H] where X is a first peak element, for the (up to) 11 ``LP first stars'' studied here (black dots). The blue open diamond represents the peculiar C-rich and Ba-poor star CS\,22949-037. The two reference stars HD\,122563 (r-poor star) and CS\,31082-001 (r-rich star) are  represented by an open and a filled red star symbol, respectively. The dashed line shows the one-to-one correspondence line.  The typical error is indicated in the right corner of the figures.}
\label{firstH}
\end{center}
\end{figure*}

% Figure 3
\begin{figure*}
\begin{center}
\resizebox{5.6cm}{5.6cm}
{\includegraphics {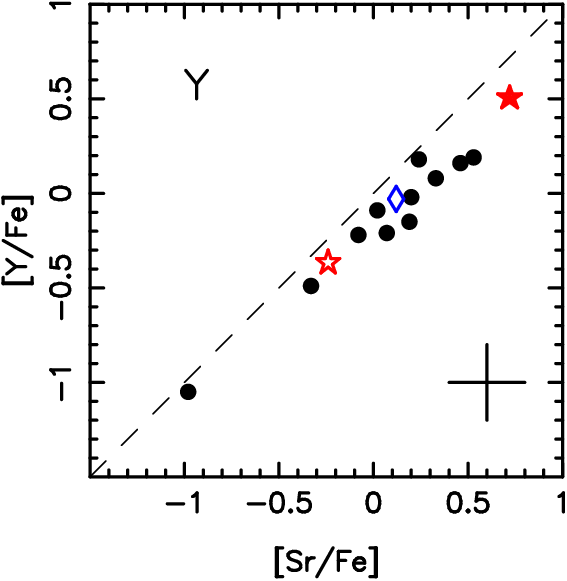}}
\resizebox{5.6cm}{5.6cm}  
{\includegraphics {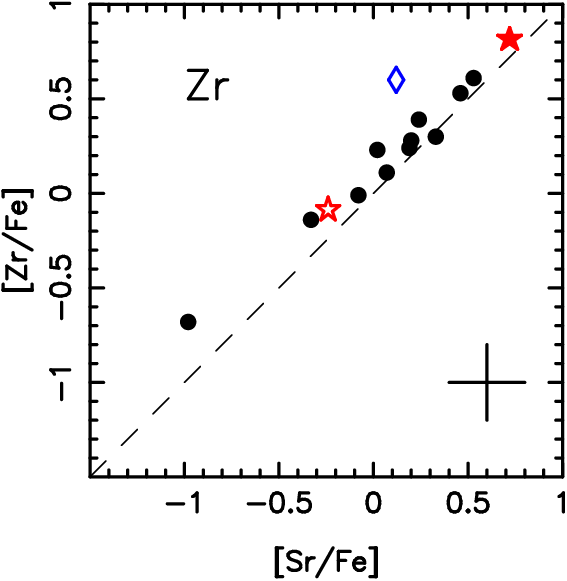}}
\resizebox{5.6cm}{5.6cm}  
{\includegraphics {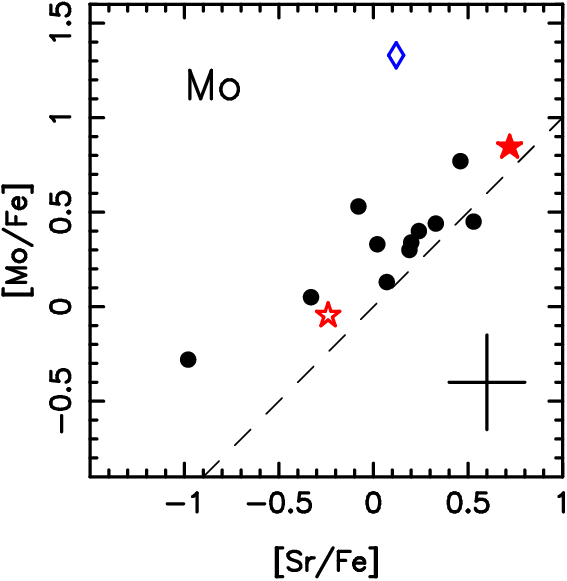}}
\resizebox{5.6cm}{5.6cm}  
{\includegraphics {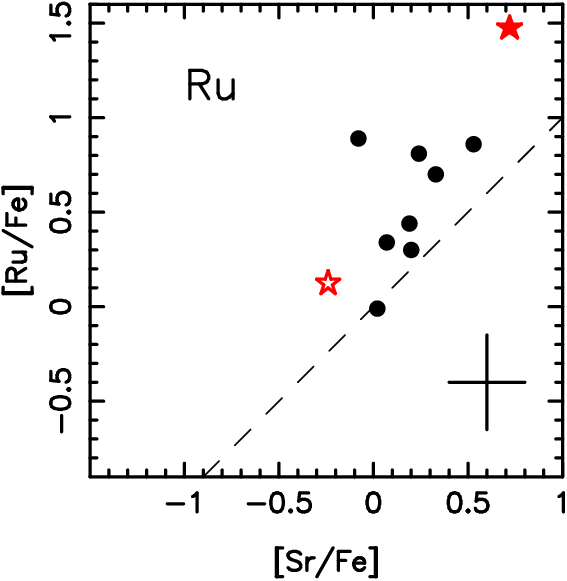}}
\resizebox{5.6cm}{5.6cm}  
{\includegraphics {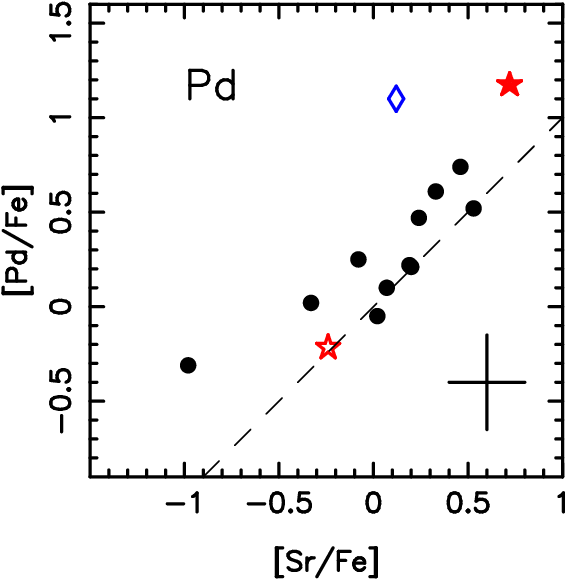}}
\resizebox{5.6cm}{5.6cm}  
{\includegraphics {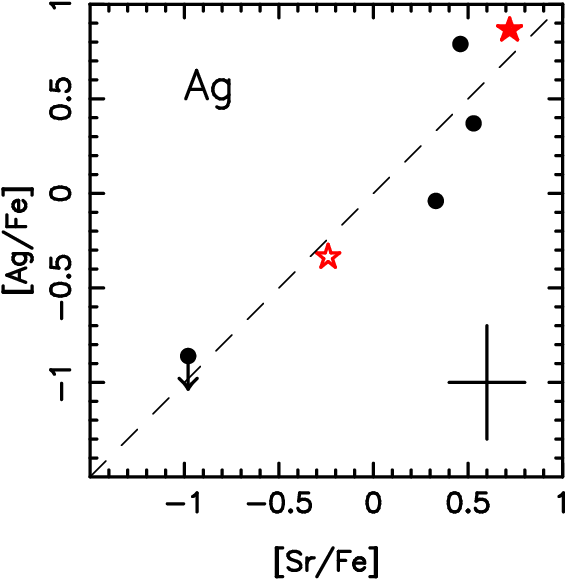}}
\caption[]{Same as Fig. \ref{firstH}, but for [X/Fe] vs. [Sr/Fe].}
\label{firstFe}
\end{center}
\end{figure*}

Generally, in VMP and EMP stars, only the abundances of the first peak elements Sr, Y, and Zr appear in the literature because the rather strong spectral lines of these elements are observable in the visible domain; there are few measurements of the abundance of the heavier first peak elements Mo, Ru, Pd, and Ag because the absorption lines of these elements are located in the less accessible near UV domain \citep[see][]{HansenP11,HansenPH12,HansenAC14,AokiIA17}. 
In the 12 giant stars we studied, it was possible to measure the abundance of these elements for the first
time because the  spectrograph UVES is very efficient in the near UV and  the S/N of the spectra, even in this region, was rather good. In Table \ref{sn} we give the S/N of the spectra measured at 385\,nm. Because of line crowding, it is not possible to measure the S/N at lower wavelengths  with precision,  but from the  ESO exposure time calculator of UVES, for the kind of stars here studied, the S/N is about four times less at 340\,nm.\\

\section {Abundance analysis}

A classical analysis of the spectra was carried out using OSMARCS model atmospheres \citep[e.g.][]{GustafssonBE75,GustafssonEE03,GustafssonEE08}.  The synthetic profiles were computed using the local thermodynamic equilibrium (LTE) spectral line analysis code {\tt turbospectrum} \citep{AlvarezP98,Plez12}. The atmospheric parameters (Table \ref{sn}) were taken from
\citet{CayrelDS04}. 
Examples of fits of neutron capture element profiles are shown in Fig. \ref{fits}. 

In Tables \ref{abund1} and \ref{abund2} we list the abundances of the neutron-capture elements of the first peak with the abundances of Ba and Eu taken from \citet{FrancoisDH07}. 
One of these stars, CS\,22949-037 \citep{DepagneHS02}, is a peculiar star; it is EMP, but C-rich and r-poor. The abundances of Eu and Sm given in \citet{DepagneHS02} are only upper limits, and Ba is the only second-peak element whose abundance could be measured in this star. 

%We consider that the uncertainty on these values is close to $\rm \pm 0.2~dex$, mainly due to the uncertainty in the position of the continuum (see Fig. \ref{fits}).

The absorption line of Ag  at 338.2\,nm is severely blended in the spectrum by a line of the NH molecule at practically the same wavelength. This NH line is relatively strong in the mixed (evolved) giants, where a mixing occurs between the atmosphere and the H-burning layer where C is converted into N by the CNO cycle. In order to have the best representation of the NH band in this region, it is necessary to use the abundance of nitrogen deduced from the NH band and not the CN band, since the abundance of N  (deduced from the CN molecule) is, in these stars, systematically lower by about 0.4\,dex than the abundance deduced from the NH molecule \citep{SpiteCP05}. In any case, the abundance of Ag is very uncertain in such N-rich stars where the contribution of Ag to the absorption feature is small (Fig. \ref{fits}, bottom panel) .

In Table \ref{abund2} we give the abundances of these elements relative to iron. The solar abundances adopted for these computations are from \citet{LoddersPG09}. In \citet{FrancoisDH07} the solar abundances of \citet{GrevesseSauval00} were adopted; {\bf  the small difference in the adopted solar abundances
%\LEt{Please specify what this refers to.}
mainly explains the small differences in the [X/Fe] values between the tables of \citet{FrancoisDH07} and the values given in Table~\ref{abund2}}. 
In some rare cases, additional measurements were carried out because new spectra became available, enabling an increase in the S/N. 
At the end of Tables \ref{abund1} and \ref{abund2}, we give, as a comparison, the abundances of the heavy elements in the two extreme  stars CS\,31082-001 \citep{SiqueiraSB13} and in HD\,122563 \citep{HondaAI07}. These stars have almost the same metallicity (-2.9 and -2.8, respectively), but the [Ba/Fe] and [Sr/Ba] ratios are very different. 

\subsection{Comparison of our measurements with previous results}
No stars in our sample are in common with the sample of \citet{AokiIA17} and only one star is in common with those of \citet{HansenPH12,HansenAC14}: HD186478. With a metallicity [Fe/H]=--2.59 this star is one of the less metal-poor stars of our sample. 
Our abundances for HD\,186478 agree with the values given by \citet{HansenPH12,HansenAC14} within the errors, except Ag. 
HD\,186478 is a cool giant star with an important mixing between the H burning layer and the surface, and the abundance of nitrogen deduced from the NH band reaches $\rm[N/Fe] \approx +1.0$ in this star. With this high value of the nitrogen abundance, the NH molecule is able to explain fully the feature at 338.3\,nm (see Fig.\ref{fits}, bottom panel for another mixed star). Generally speaking we decided not to take into account the abundance of Ag in the mixed giants.

% Figure 4
\begin{figure*}
\begin{center}
\resizebox{7.2cm}{8.6cm}                  
{\includegraphics {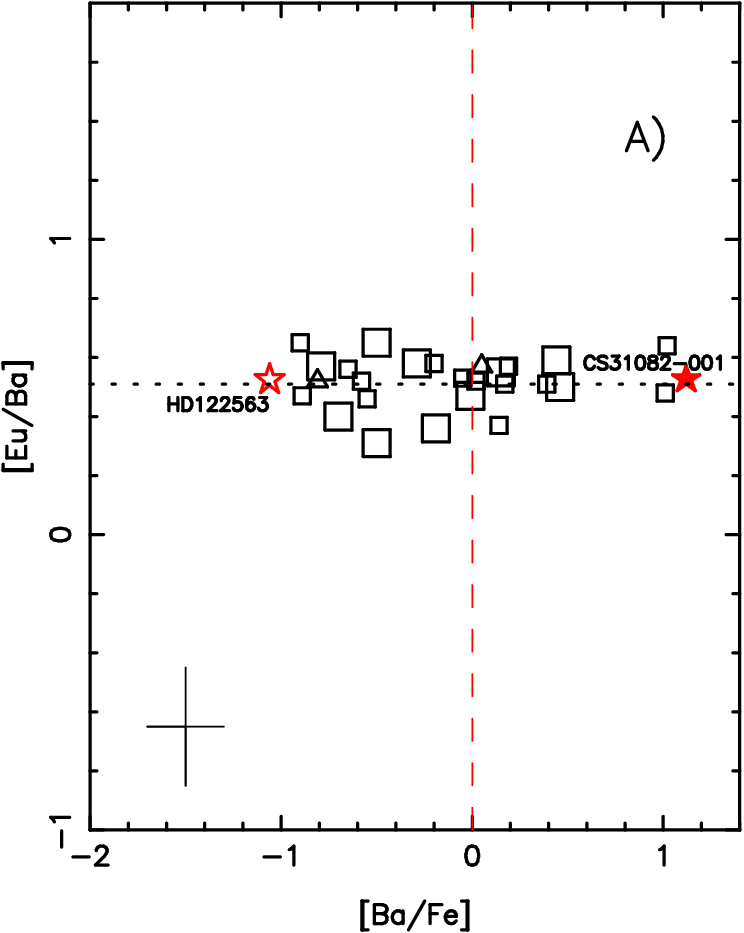}}
\resizebox{7.2cm}{8.6cm}                  
{\includegraphics {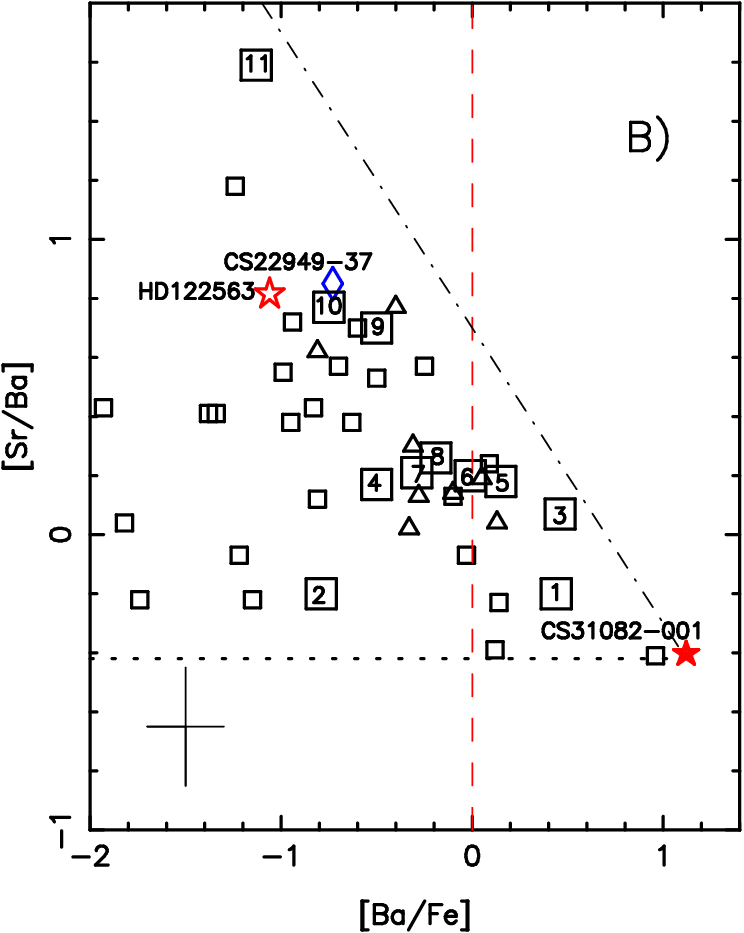}}
\caption[]{[Eu/Ba] and [Sr/Ba] vs. [Ba/Fe] for the stars studied from the LP First Stars and the stars studied by  \citet{SiqueiraHB14} and \citet{SiqueiraAB15}.
 The open triangles represent the turn-off stars and the open squares the giants. The big square symbols indicate the 11 giants studied here and the number given at the centre of the squares in panel B refers to the identification of the star in Tables  \ref{abund1} and \ref{abund2} (with higher numbers for higher [Sr/Ba]). CS\,31082-001 (red filled star symbol) and HD\,122563 (red open star symbol) are the `=reference stars. CS\,22949-037 (blue open diamond, only on panel B) is a peculiar C-rich metal-poor giant. 
 
In panels A and B a red dashed line is drawn at [Ba/Fe]=0.0; in panel A the dotted line represents the mean LTE value of [Eu/Ba] in VMP stars (close to  a pure r process value as observed in CS\,31082-001) and in panel B the minimum of [Sr/Ba] close to a pure r-process production as observed in CS\,31082-001. The dash-dotted line in panel B indicates the upper limit of [Sr/Ba].}
\label{basr}
\end{center}
\end{figure*}

%TABLE 2
\begin{table*}
\begin{center}   
\caption[]{
Metallicities [Fe/H], and $\rm A(X) = log(N_{X} / N_{H}) + 12$ of the elements for the 7 neutron-capture elements of the first peak (+  A(Ba) and A(Eu) ) derived from LTE computations in our 11 ``LP first stars''. In column 1 lists the identification number ID of the star, also included  in Fig. \ref{basr}{\tiny B}. ID=1 is for the star with the lowest [Sr/Ba] ratio and ID=11 for the star with the highest [Sr/Ba] ratio (see [Sr/Ba] in Table \ref{abund2}).\\
At the end of the table, we give the abundances of these elements in the peculiar C-rich star CS\,22949-037 and in the classical r-rich and r-poor stars  CS\,31082-001 and HD~122563. }
\label{abund1}
%\begin{tabular}{l@{~~}c@{~~}c@{~~}c@{~~}c@{~~}c@{~~}c@{~~}c@{~~}c@{~~}c@{~~}c@{~~}c@{~~}c}
\begin{tabular}{clrrrrrrrrrrrrrr}
\hline
ID&  Object       &         [Fe/H]&   A(Sr)&    A(Y)&   A(Zr)&   A(Mo)&    A(Ru)&    A(Pd)&   A(Ag)&   A(Ba)&   A(Eu)\\
\\
  &Sun*           &          0.0  &    2.90&    2.20&    2.57&    1.94&     1.78&     1.67&    1.22&    2.18&    0.53\\
\\
1 &CS\,22953-003  &         --2.84&    0.30&  --0.46&    0.12&  --0.50&   --0.25&   --0.70&    --  &  --0.22&  --1.28\\
2 &BD--18\,5550   &         --3.06&  --1.15&  --1.91&  --1.17&  --1.50&  <--1.70&   --1.70& <--2.70&  --1.67&  --2.75\\
3 &CS\,22896-154  &         --2.69&    0.74&  --0.30&    0.49&  --0.30&   --0.05&   --0.50&  --1.10&  --0.05&  --1.20\\
4 &CS\,22873-055  &         --2.99&  --0.42&  --1.28&  --0.56&  --1.00&  <--1.70&   --1.30&    --  &  --1.31&  --2.65\\
5 &BS~17569-049   &         --2.88&    0.35&  --0.60&  --0.01&  --0.50&   --0.40&   --0.60&  --1.70&  --0.55&  --1.65\\
6 &HD~186478      &         --2.59&    0.50&  --0.54&    0.22&  --0.35&   --0.37&   --0.70&    --  &  --0.42&  --1.60\\
7 &CS\,22966-057  &         --2.62&    0.20&  --0.64&  --0.06&  --0.15&     0.05&   --0.70&    --  &  --0.73&  --1.80\\
8 &HD~2796        &         --2.47&    0.50&  --0.48&    0.21&  --0.40&   --0.35&   --0.70&    --  &  --0.48&  --1.77\\
9 &CS\,29518-051  &         --2.69&    0.41&  --0.51&    0.16&  --0.50&   --0.70&   --0.90&    --  &  --1.01&  --2.01\\
10&CS\,22873-166  &         --2.97&  --0.05&  --0.86&  --0.17&  --0.70&   --1.20&   --1.30&    --  &  --1.54&  --2.76\\
11&CS\,22897-008  &         --3.41&  --0.05&  --1.05&  --0.31&  --0.70&    --  &    --1.00&  --1.40&  --2.36&    --  \\
\\
  &CS\,22949-037  &         --3.97&  --0.95&  --1.80&  --0.87&  --0.70&    --  &    --1.20&    --  &  --2.52&    --  \\
\\
 &CS\,31082-001  &          --2.90&    0.72&  --0.19&    0.49&  --0.11&    0.36&    --0.05&  --0.81&    0.40&  --0.72\\
 &HD~122563      &          --2.77&  --0.11&  --0.93&  --0.28&  --0.87&  --0.86&    --1.31&  --1.88&  --1.65&  --2.77\\
\hline    
%\multicolumn{10}{l}{* ~CS\,22949-037 is a peculiar Carbon-rich metal-poor star \citep{DepagneHS02}.}
\multicolumn{10}{l}{* The solar abundances $A(X)_{\odot}$ ~are taken from \citet{LoddersPG09}.}
\end{tabular}  
\end{center}   
\end{table*}

%TABLE 3
\begin{table*}
\begin{center}    
\caption[]{ Same as Table \ref{abund1}, but with [X/Fe]. [Sr/Ba] and [Eu/Ba] are given in columns 13 and 14.
 The computations are carried out for $\rm A(Fe)_{\odot}=7.50$.
}
\label{abund2}
\begin{tabular}{c@{~~}l@{~~}r@{~~}r@{~~}r@{~~}r@{~~}r@{~~}r@{~~}r@{~~}r@{~~}r@{~~}r@{~~~~}r@{~~}r@{~~}}
%\begin{tabular}{clrrrrrrrrrrrr}
\hline
ID &Object        &  [Fe/H]& [Sr/Fe]&  [Y/Fe]& [Zr/Fe]& [Mo/Fe]& [Ru/Fe]& [Pd/Fe]& [Ag/Fe]& [Ba/Fe]& [Eu/Fe]& [Sr/Ba]& [Eu/Ba]\\
\\
1 &CS\,22953-003  &   --2.84&    0.24&    0.18&    0.39&    0.40&    0.81&    0.47&    --  &    0.44&    1.03&--0.20 & 0.59\\
2 &BD--18\,5550   &   --3.06&  --0.99&  --1.05&  --0.68&  --0.38&     -- &  --0.31&    --  &  --0.79&  --0.22&--0.20 & 0.57\\
3 &CS\,22896-154  &   --2.69&    0.53&    0.19&    0.61&    0.45&    0.86&    0.52&    0.37&    0.46&    0.96&  0.07 & 0.50\\
4 &CS\,22873-055  &   --2.99&  --0.33&  --0.49&  --0.14&    0.05&     -- &    0.02&    --  &  --0.50&  --0.19&  0.17 & 0.31\\
5 &BS~17569-049   &   --2.88&    0.33&    0.08&    0.30&    0.44&    0.70&    0.61&  --0.04&    0.15&    0.70&  0.18 & 0.55\\
6 &HD~186478      &   --2.59&    0.19&  --0.15&    0.24&    0.30&    0.44&    0.22&    --  &  --0.01&    0.46&  0.20 & 0.47\\
7 &CS\,22966-057  &   --2.62&  --0.08&  --0.22&  --0.01&    0.53&    0.89&    0.25&    --  &  --0.29&    0.29&  0.21 & 0.58\\
8 &HD~2796        &   --2.47&    0.07&  --0.21&    0.11&    0.13&    0.34&    0.10&    --  &  --0.19&    0.17&  0.26 & 0.36\\
9 &CS\,29518-051  &   --2.69&    0.20&  --0.02&    0.28&    0.25&    0.21&    0.12&    --  &  --0.50&    0.15&  0.70 & 0.65\\
10&CS\,22873-166  &   --2.97&    0.02&  --0.09&    0.23&    0.33&  --0.01&    0.00&    --  &  --0.75&  --0.32&  0.77 & 0.43\\
11&CS\,22897-008  &   --3.41&    0.46&    0.16&    0.53&    0.77&    --  &    0.74&    0.79&  --1.13&    --  &  1.59 &  -- \\
\\
  &CS\,22949-037  &   --3.97&    0.12&  --0.03&    0.53&    1.33&    --  &    1.10&    --  &  --0.73&    --  &  0.85 &  -- \\
\\
&CS\,31082-001    &   --2.90&    0.72&    0.51&    0.82&    0.85&    1.48&    1.18&    0.87&    1.12&    1.65& -0.43 & 0.53\\
&HD~122563        &   --2.77&  --0.24&  --0.36&  --0.08&  --0.04&    0.13&  --0.21&  --0.33&  --1.06&  --0.53&  0.82 & 0.53\\
\hline    
%\multicolumn{10}{l}{* ~CS\,22949-037 is a peculiar Carbon-rich metal-poor star \citep{DepagneHS02}.}
\end{tabular}
\end{center} 
\end{table*}

% Figure 5
\begin{figure*}
\begin{center}
\resizebox{6cm}{3.5cm}                  
{\includegraphics {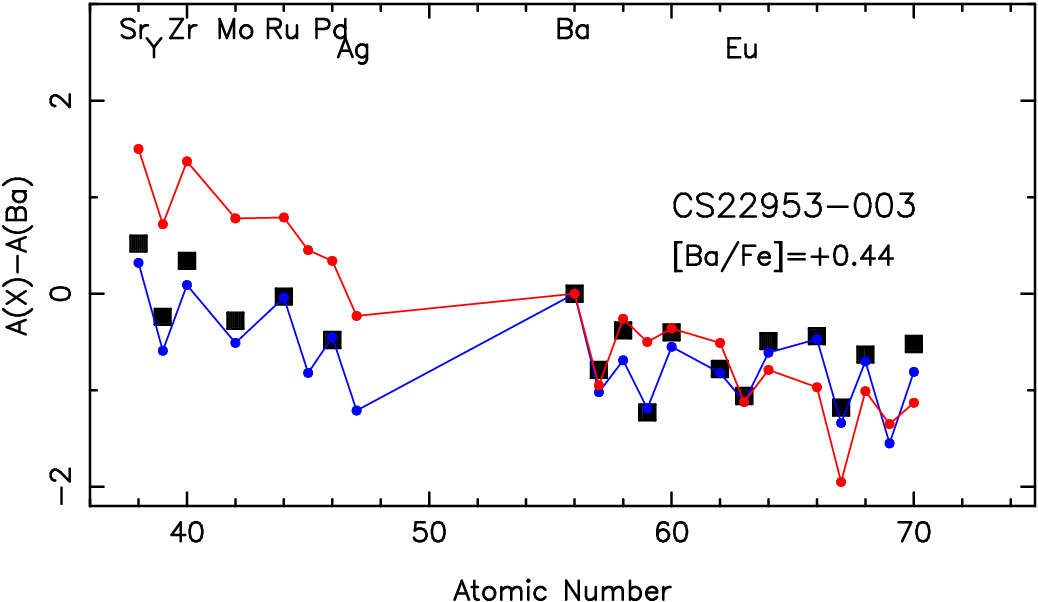}}
\resizebox{6cm}{3.5cm}  
{\includegraphics {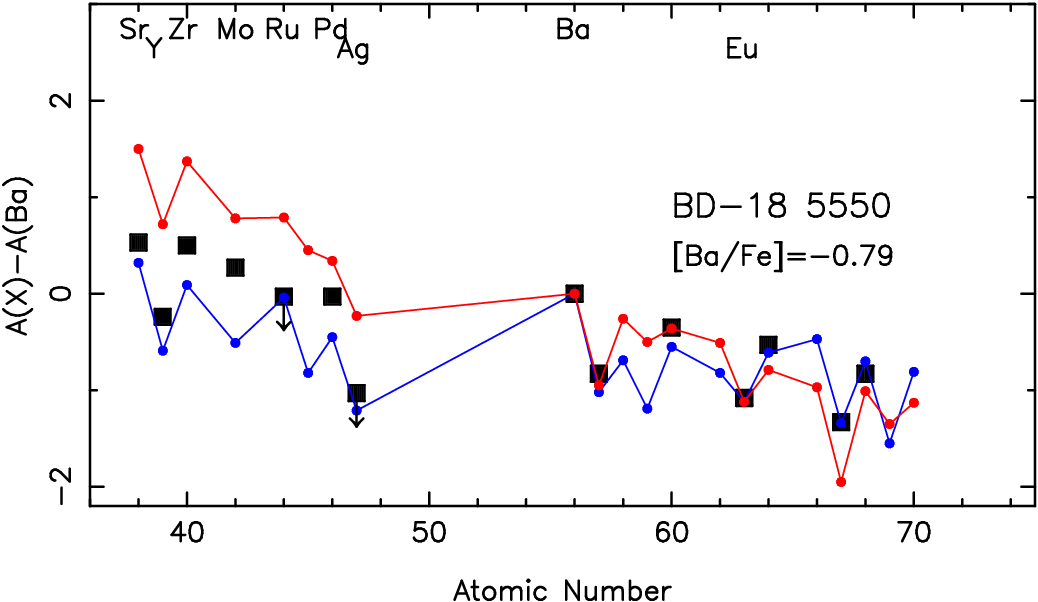}}
\resizebox{6cm}{3.5cm}  
{\includegraphics {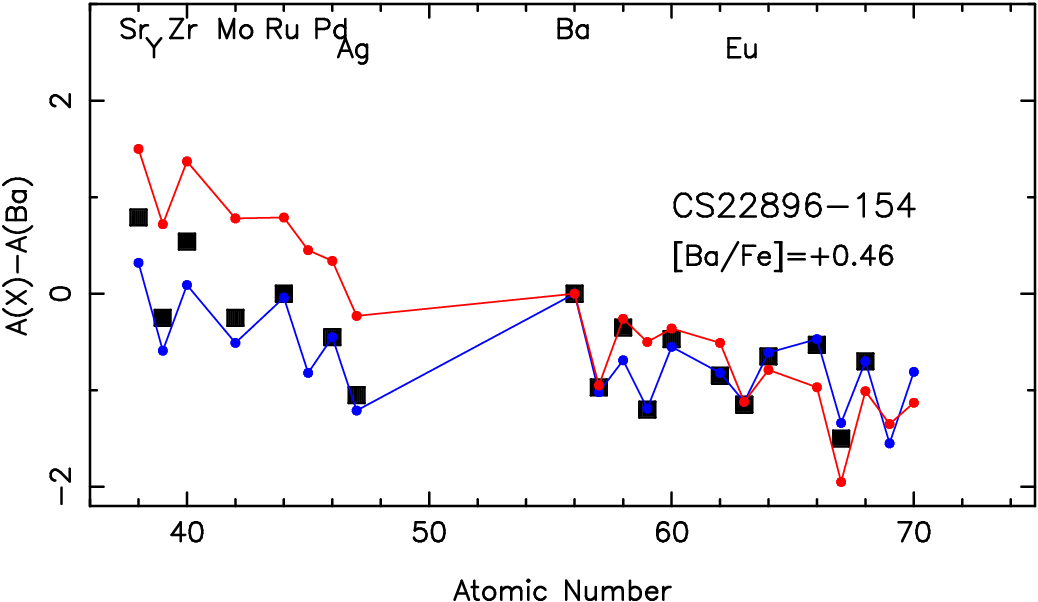}}
\caption[]{Abundance patterns of the neutron capture elements for the stars of our sample with $\rm[Sr/Ba]<0.10$. The blue line represents the pattern in CS\,31082-001, which is a standard r-rich star, and the red line represents the pattern in HD\,122563, which is an r-poor star with a high value of [Sr/Ba].}
\label{pat1}
\end{center}
\end{figure*}

% Figure 6
\begin{figure*}
\begin{center}
\resizebox{6cm}{3.5cm}  
{\includegraphics {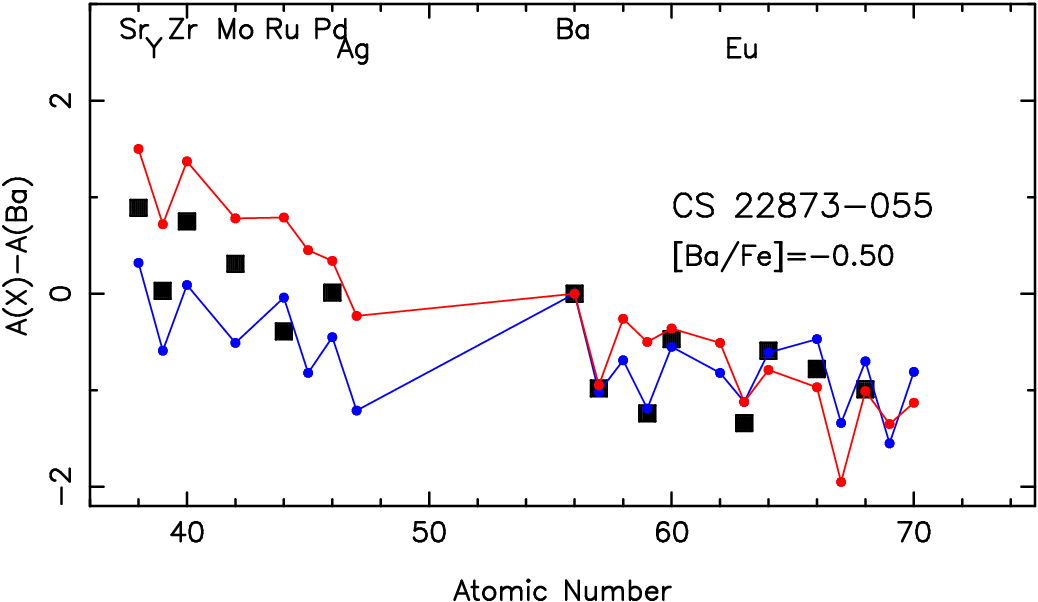}}
\resizebox{6cm}{3.5cm}  
{\includegraphics {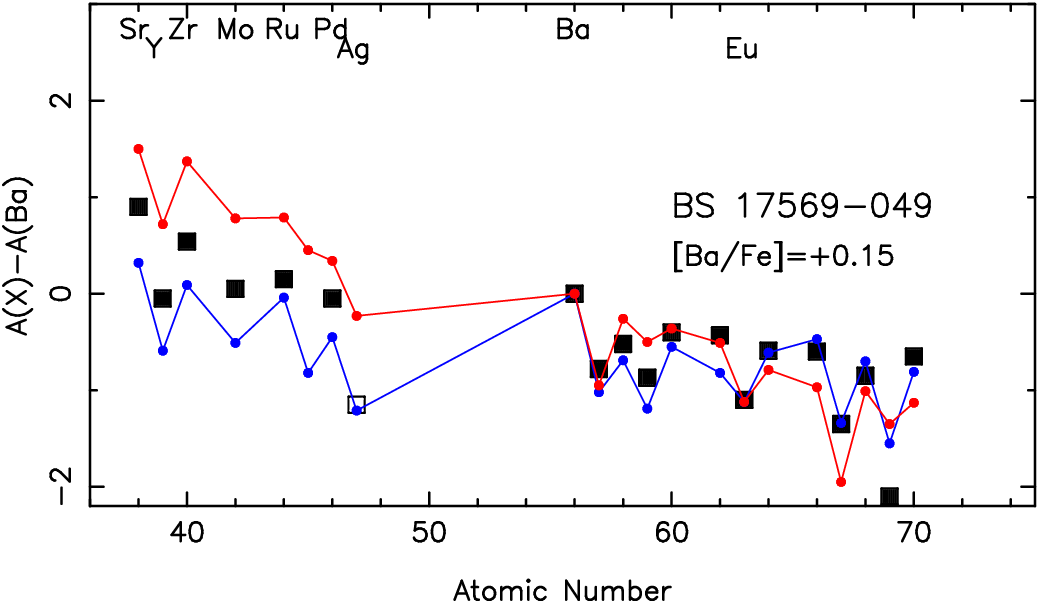}}
\resizebox{6cm}{3.5cm}  
{\includegraphics {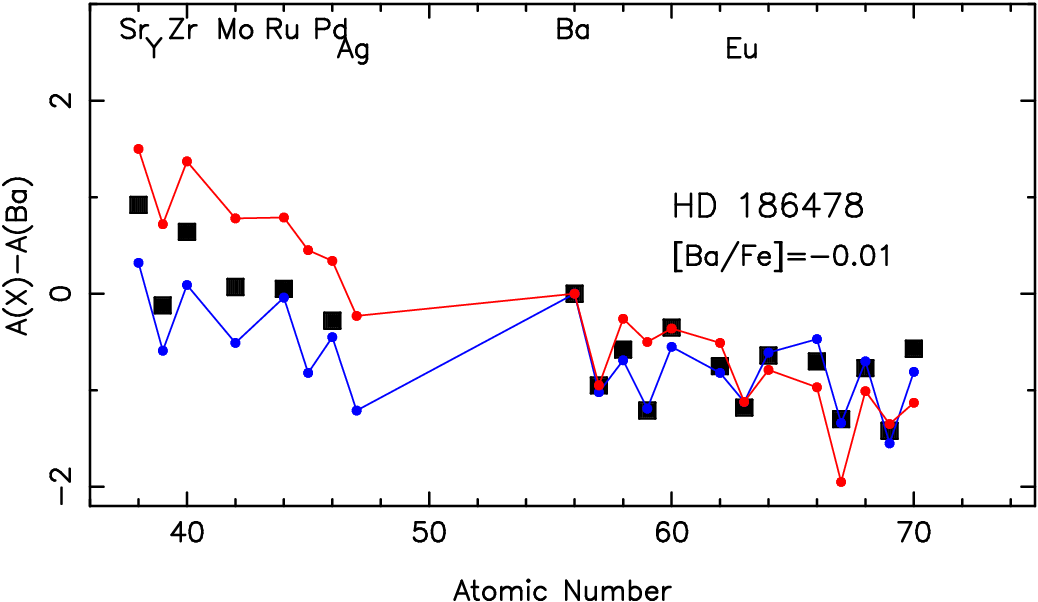}}
\resizebox{6cm}{3.5cm}  
{\includegraphics {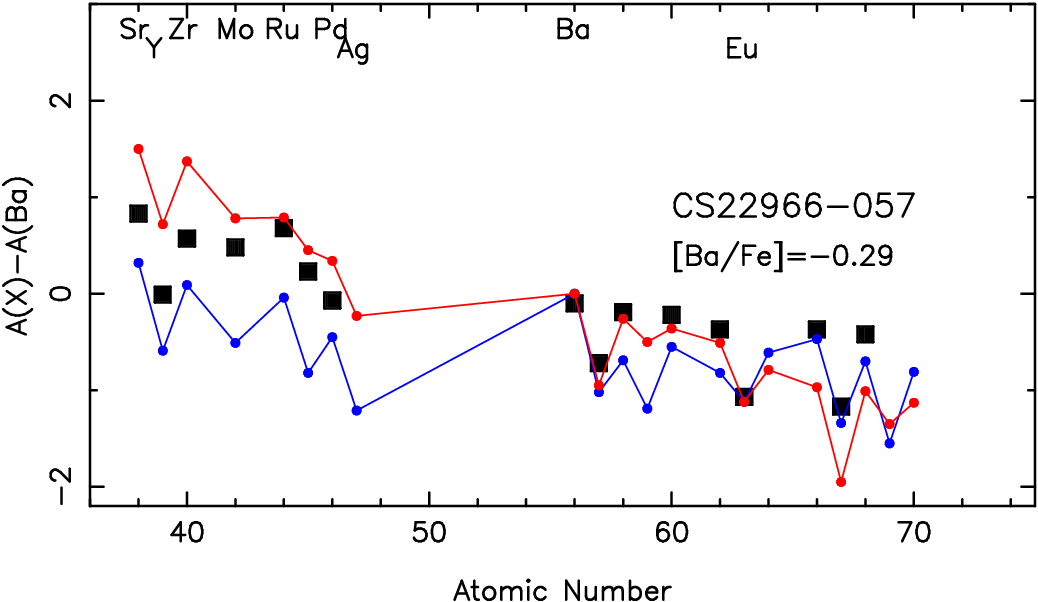}}
\resizebox{6cm}{3.5cm}  
{\includegraphics {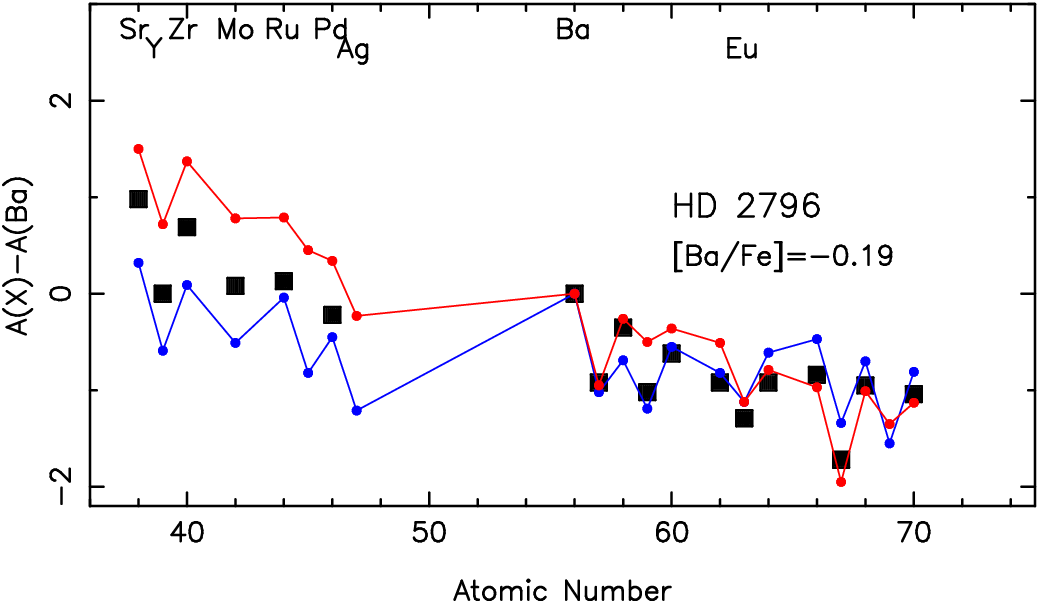}}
\caption[]{Abundance patterns of the neutron capture elements for the stars of our sample with the intermediate ratios: $\rm 0.10<[Sr/Ba]<0.50$. The symbols are the same as in Fig.\ref{pat1}. }
\label{pat2}
\end{center}
\end{figure*}

% Figure 7
\begin{figure*}
\begin{center}
\resizebox{6cm}{3.5cm}
{\includegraphics {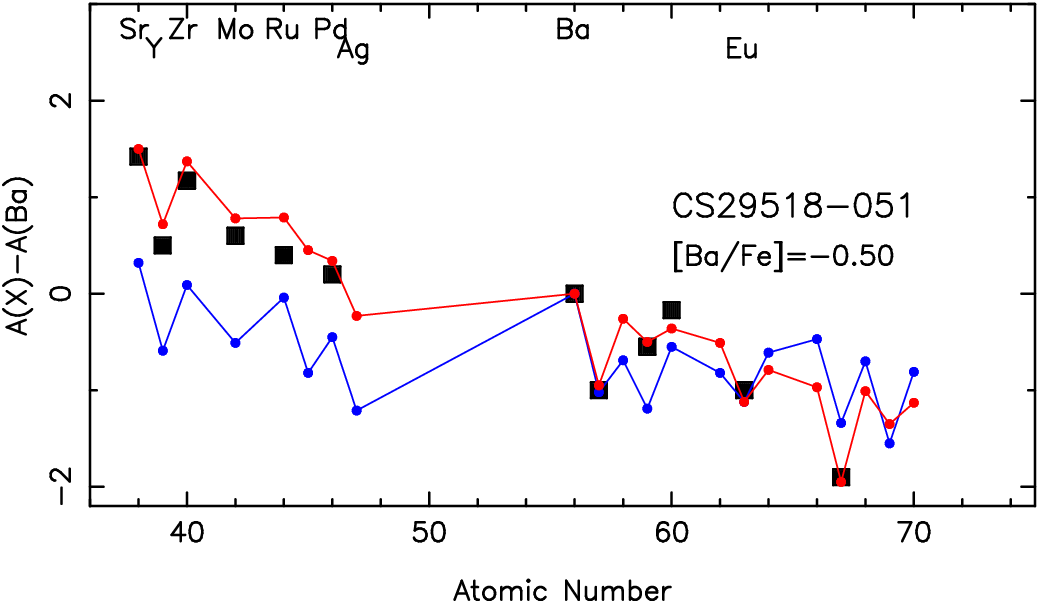}}
\resizebox{6cm}{3.5cm}  
{\includegraphics {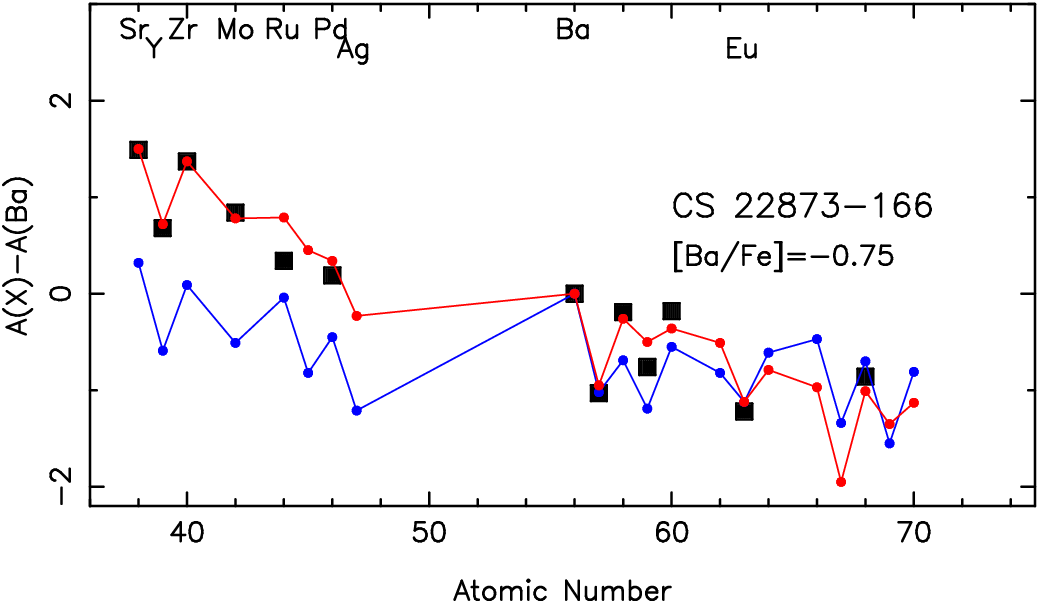}}
\caption[]{Abundance patterns of the neutron capture elements for the two stars of our sample very similar to HD\,122563 (with $\rm [Sr/Ba]\sim0.75$). The symbols are the same as in Fig.\ref{pat1}.}
\label{pat3}
\end{center}
\end{figure*}

% Figure 8
\begin{figure}
\begin{center}
\resizebox{7.5cm}{4.4cm}  
{\includegraphics {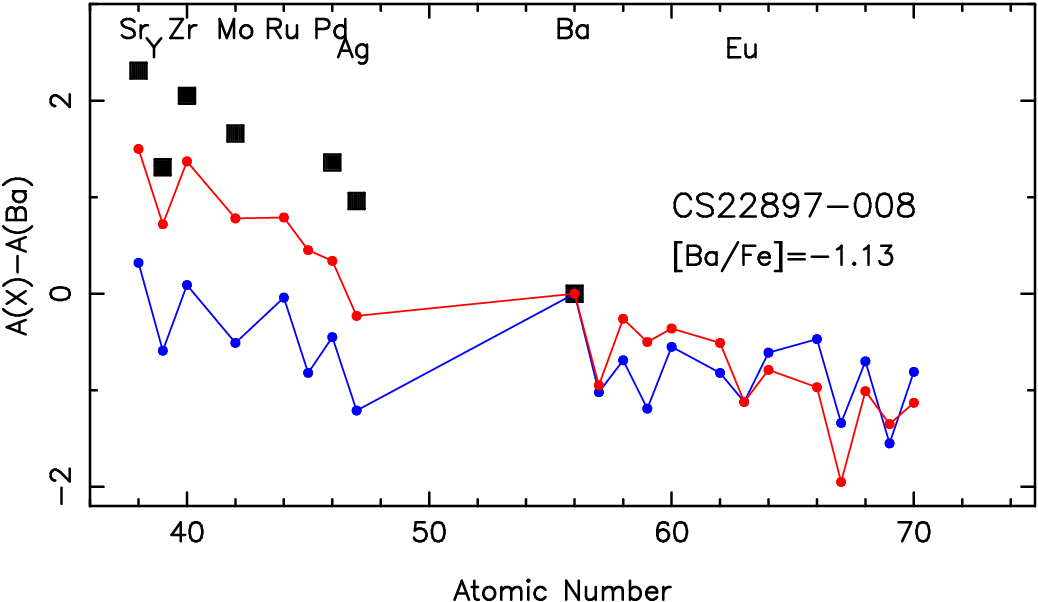}}
\caption[]{Abundance pattern of the neutron capture elements in CS~22897-008 (with $\rm [Sr/Ba]\sim1.6$), compared to CS\,31082-001 (blue line) and HD\,122563 (red line).}
\label{pat4}
\end{center}
\end{figure}

\subsection{Error budget}

The error in [X/Fe] linked to our choice of the model parameters was computed
as a quadratic sum of the uncertainties on \Teff ~(100K), \logg~(0.2 dex) and the turbulence velocity (\vt = 2 \kms). It is less than 0.08\,dex. It is small compared to the error made in fitting the synthetic line profile to the observed spectra. The global error depends on the element (uncertainty in the position of the continuum) and was estimated to be between 0.20 and 0.25\,dex by \citet{FrancoisDH07}. We adopted err=0.2 dex for Sr, Y, and Zr, 0.25 dex for Mo, Ru, and  Pd, and  0.3\,dex for Ag.

\section{Analysis of the results}  

Among the EMP and VMP stars, there is a very large scatter of neutron-capture element abundances. Some are europium-rich and other europium-poor. Since Eu can be formed almost only by the r-process, these stars are classified as r-rich or r-poor according to their [Eu/Fe] value.
It has been shown (but often overlooked) that, in the old stars, there is an excellent correlation between the two ratios [Eu/Fe] and [Ba/Fe]. In Fig.\ref{basr}{\tiny A}  [Eu/Ba] is plotted versus [Ba/Fe] for the stars studied from the  LP First Stars and the stars studied in a similar way by \citet{SiqueiraHB14} and \citet{SiqueiraAB15}. In these very old stars,  we found $\rm[Eu/Ba] =0.51 \pm 0.09$ dex. The constancy of the ratio [Eu/Ba] has also been noted by \citet{MashonkinaCB10}. In their figure 12, at low metallicity, below [Fe/H]=--2.5,  they  found  $\rm A(Ba)-A(Eu) \approx 1.1 dex$ or $\rm[Eu/Ba] \approx 0.56$ dex in good agreement with our mean value. This value is close to the pure r-process value $\rm[Eu/Ba] \approx 0.7$ dex \citep{BanerjeeQH17}.

In such old stars, europium-rich stars are also Ba-rich (and conversely). As a consequence, at very low metallicity since, in particular in turn-off stars and most metal-poor giants, the Eu lines are extremely weak and a very high resolution and S/N ratio are required to measure these lines  \citep[see for example][]{SiqueiraBS12}, it is possible to use Ba as a proxy of Eu.  From  Fig. \ref{basr}{\tiny A}, it can be deduced that in normal VMP  stars $\rm[Eu/Fe] = [Ba/Fe] + 0.51 \pm0.09$ (or  $\rm[Eu/H] = [Ba/H] + 0.51$). This is of course generally not valid for carbon-rich metal-poor stars enriched in neutron-capture elements by mass transfer by a more massive companion during its AGB phase.

In the next sections, we compare the relative abundances of the neutron-capture elements in the  11 LP  First Stars with the abundances measured in two classical stars: that is, CS~31082-001, the well-known r-rich star \citep{HillPC02,SiqueiraSB13}; and HD~122563 \citep{HondaAI07}, the well-known r-poor star.
 
\subsection{Correlations between the abundances of the first peak elements}
Generally speaking, in all the EMP and VMP stars there is a rather good correlation between the abundance
patterns of the second peak elements, extended from Ba to Yb, \citep[see e.g.][]{FrancoisDH07,MontesBC07,SiqueiraHB14}.
We wonder about the abundances of the first peak (from Sr to Ag).

In Fig. \ref{firstH} we plot [X/H] versus [Sr/H], where X is for Y, Zr, Mo, Ru, Pd, and Ag, and in Fig. \ref{firstFe} [X/Fe] versus [Sr/Fe]. 
The excellent correlation between [Y/Fe] and [Sr/Fe] was already underlined by \citet{FrancoisDH07}. For the other elements of the first peak, the correlation is also good but the scatter is larger \citep[see also][]{HansenPH12}. 

\subsection{A peculiar object: CS~22949-37}
The star CS~22949-37 \citep{DepagneHS02,NorrisRB02} is classified as CEMP-no by \citet{MasseronJP10}, i.e. C-rich and EMP, without enhancement of the neutron-capture elements. The abundance anomalies in C-rich stars are often explained by recent pollution by a companion in its AGB phase. But the radial velocity $\rm V_{R}$ of CS~22949-37 does not seem to be variable; in 2000 and 2001 \citet{DepagneHS02} found $\rm V_{R}= -125.64$ and $-125.62\pm0.2$ \kms and in 2013 \citet{StarkenburgSM14} measured $\rm V_{R}=-125.9 \pm0.3$ \kms. Therefore it is improbable that the star be a binary unless it is seen pole on. The abundances of this star should represent the abundances in the cloud that formed the star.   \citet{ChoplinEM17}  have explained the observed distribution of the elements (from C to Fe) in single CEMP stars  by the elements produced by fast rotating massive stars (FRMS). With its high overabundance of Na and Mg and with a ratio $\rm^{12}C/^{13}C=4.0$, the element pattern of CS~22949-37, is rather similar to the pattern of HE\,2139-5432 fairly well represented by the model 20s7mix \citep[figure 11 of][]{ChoplinEM17}. 

In Fig. \ref{firstFe}, the star CS~22949-37 has a normal [Y/Fe] ratio compared to the other LP first stars but it seems to be rich in Zr, Mo, and Pd. The very weak lines of Ru could not be measured in this EMP star with [Fe/H]=--4. 

%(Since the star is a cool giant, it is possible that part of the carbon in the atmosphere of the star was transformed into nitrogen because of an internal mixing with the CNO burning layer, but the neutron-capture elements are not affected by this process.)

\subsection{Sr/Ba scatter in metal-poor stars}
The abundance of the elements inside a given peak (first or second) are well correlated (see Fig. \ref{firstFe}). 
But there is no clear correlation between the abundance of a first peak element (like Sr) and the abundance of a second peak element (like Ba).
For example in Fig. \ref{basr}{\tiny B} we plot the [Sr/Ba] ratio versus [Ba/Fe] for all the stars of the ESO LP First Stars and the stars studied in a homogeneous way by \citep{SiqueiraBS12,SiqueiraAB15}. 
The position of the 11 EMP stars analysed here, are indicated, by numbered open squares. These stars are identified by numbers from 1 (for the star with the lowest [Sr/Ba] ratio) to 11 for the star with the highest ratio (see Table 2). The position of the CEMP-no star CS~22949-37 is indicated with a blue diamond. 
All the stars in this figure  have a metallicity [Fe/H] below or equal to --2.5,  of which most are concentrated in the rather narrow range $\rm [Fe/H]=-3.1\pm 0.3$ dex. 
%  The stars outside this small interval of metallicity are indicated by a cross on the figure. 

\citet{Roederer13} remarked that in field stars the scatter of the relation [Ba/H] versus [Sr/H] increases when [Ba/H] decreases. It is  clear from Fig. \ref{basr}{\tiny B} that the scatter of [Sr/Ba] strongly increases when [Ba/Fe] decreases \citep{SpiteS14,SpiteSB14,JacobsonKF15}.
The horizontal dotted line in Fig. \ref{basr}{\tiny B} represents the minimum of [Sr/Ba] corresponding to its value in the r-rich stars and characteristic of the main r-process. The dash-dotted line which, for each value of [Ba/Fe], seems to mark the upper limit of [Sr/Ba], corresponds to the line at [Sr/Ba]=--[Ba/Fe] + 0.7. The existence of this limit means in fact, that [Sr/Fe] is never higher than about +0.7\,dex a value close to the value observed (at this metallicity) in the r-rich stars, such as CS\,31082-001, CS\,22892-052 \citep{SnedenCL03,SnedenCG08}, 
CS\,29497-004 \citep{2016HillCB},
or RAVE J203843.2--002333 \citep{PlaccoHF17}.
This is also the observed upper limit of [Sr/Fe] at all metallicities following \citet[][see their Fig.1]{FrancoisDH07}.

The stars with a ratio [Ba/Fe] around \,--1.\,dex, show a range of [Sr/Ba] spanning about 2\,dex. For example, the star BD--18\,5550 (labelled 2 in Fig.\ref{basr}{\tiny B}) has about the same Sr/Ba ratio as the classical r-rich star (CS\,31082-001), but one of our stars (CS 22897-008, labelled 11 in Fig.\ref{basr}{\tiny B}) has a  Sr/Ba ratio that is even higher than the ratio measured in the extreme star HD\,122563.\\  
Below [Ba/Fe]=--1, the scatter of [Sr/Ba] does not continue to increase (it even seems to decrease) but the number of stars is very small and this partial trend is probably not significant.

% Figure 9
\begin{figure}
\resizebox{8.4cm}{4.0cm}                   
{\includegraphics {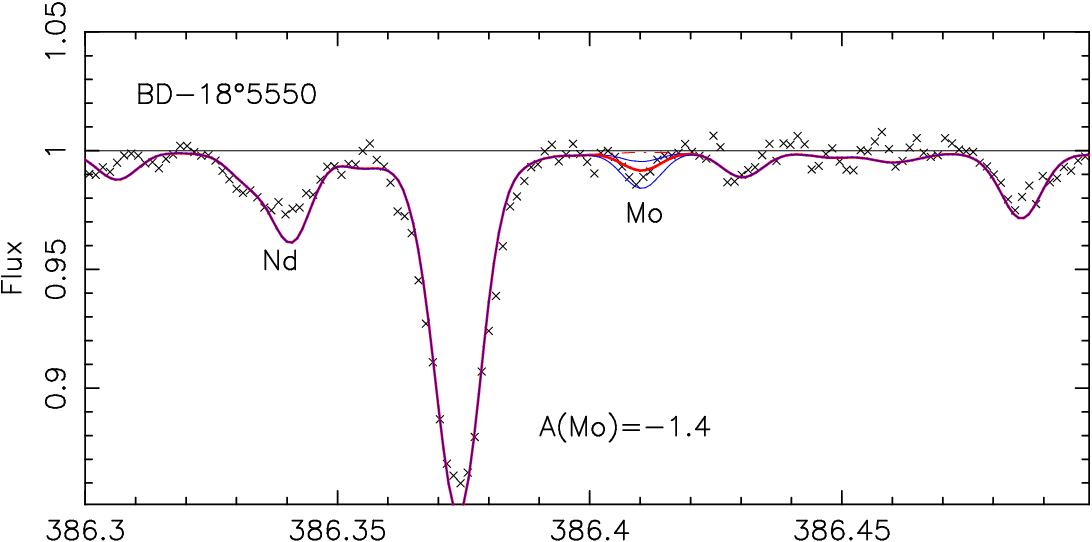}}
\caption[]{Line of Mo in BD--18\,5550. Crosses represent the observations. The synthetic profiles were computed with an absence of Mo (red dash-dotted line, here practically at the position of the continuum), the adopted abundance A(Mo)=--1.4 (red thick line), and changing this abundance by $\pm 0.3$ dex (blue thin lines for A(Mo)=--1.7 and --1.1). }
\label {Mo18}
\end{figure}

\subsection{Abundance patterns of the neutron-capture elements in our sample of stars}

In Figs.\,\ref{pat1}, ~\ref{pat2}, \ref{pat3}, and \ref{pat4} we present the abundance patterns of the neutron-capture elements in our sample of stars. The figures are sorted by increasing [Sr/Ba]. 
The abundances of the first peak elements are from Table \ref{abund1}, and the abundances of the second peak elements are from \citet{FrancoisDH07} rescaled to the solar abundances of \citet{LoddersPG09}.
The data of the first and second peak of the reference stars were taken from \citet{SiqueiraSB13} for CS\,31082-001 (blue line in the figures) and from \citet{HondaAI07} for HD\,122563 (red line). 
We did not find the Holmium (Z=67) abundance in HD\,122563 in the literature; therefore, to allow a comparison with the stars here studied (Figs.\,\ref{pat1} to \ref{pat4}), we measured the 345.6 nm line of Ho in the ESO spectra of this star and found A(Ho)=--3.6, i.e. [Ho/Fe]=--1.32 \citep[atomic data from][]{LawlerSC04}.

\subsubsection{Stars with a [Sr/Ba] ratio close to the low ratio observed in r-rich stars}
In this class we studied three stars with  $\rm[Sr/Ba]<+0.1$  (Fig.\ref{pat1}). Two of these stars, CS\,22953-003 and CS\,22896-154, for which [Ba/Fe]=+0.44 and +0.46 dex, can be considered as r-rich but the third star, BD--18\,5550, for which [Ba/Fe]=--0.79 (and [Eu/Fe]=--0.22\,dex), is an r-poor star. 
In these three stars, the abundance pattern of the first peak elements is rather close to the pattern of the typical r-rich star CS\,31082-001. In a recent paper, \citet{Roederer17} found also that the pattern of the n-capture elements in BD--18\,5550 is close to the pattern of the main r-process. However we note that, in BD--18\,5550, the abundance of molybdenum seems to be stronger than expected. In the region of the molybdenum line, the S/N ratio of the spectrum is close to 350 (Fig. \ref{Mo18}) and we derived $\rm A(Mo) = -1.4 \pm 0.2$.  This corresponds to 0.04 atoms of molybdenum   for $10^{12}$ H atoms, which is four times more than the predictions of a pure r-process. At this very low abundance of the neutron capture elements ($\rm[Ba/H]\approx -3.8$) a very weak pollution is able to alter the abundance ratios.\\

Like BD--18\,5550, three other stars studied in the LP First Stars associate  a low [Ba/Fe] ratio (below -1) and a low [Sr/Ba] ratio \citep{FrancoisDH07}. These stars are CS\,22172-002, CS\,29502-042, and CS\,22885-096. We could not study the pattern of the first peak abundances in these stars because the S/N of the obtained spectra did not allow the measurements of extremely weak lines.

\subsubsection{Stars with an `intermediate [Sr/Ba] ratio}
The five stars (Fig. \ref{pat2}) of our sample with a value of [Sr/Ba] higher than in CS\,31082-001, but lower than in HD\,122563,  have a pattern of the neutron-capture elements intermediate between these two stars. This enhancement seems to be almost uniform for all the first peak elements that we measured. However in CS\,22966-057 molybdenum seems to be more enhanced than the other first peak elements.

\subsubsection{Stars with a very high [Sr/Ba] ratio}
In Fig. \ref{pat3} and \ref{pat4} the stars have a [Sr/Ba] ratio higher than 0.7\,dex. All these stars are r-poor with  a low ratio  $\rm[Ba/Fe]\leq -0.5$.  The two stars CS\,29518-051 and CS\,22873-166 (Fig. \ref{pat3}) have the same distribution of the neutron-capture elements as HD\,122563; their first peak elements are enhanced by a factor of about ten relative to the classical r-rich star CS\,31082-001.\\ 
One of the stars analysed here, CS\,22897-008 (Fig. \ref{pat4}), seems to be even more extreme than HD\,122563; compared to the classical r-rich star CS\,31082-001, the first peak elements are almost uniformly enhanced by a factor of 100. This extreme star has a very low [Ba/H] ratio of --4.6. The lines of the heaviest elements are not visible, and the estimated values of the corresponding equivalent widths are of the order of, or less than 1m{\rm \AA}, which is obviously too small to be measured on our spectra.

\section{Discussion}
\subsection{Heavy elements enrichment in VMP and EMP stars}
\citet{Roederer13} noted that the yields of the heavy elements have to be decoupled from the production of magnesium and iron in order to explain the small observed dispersion in the ratios of light metals (such as [Mg/Fe]) and the large dispersion observed for [Eu/Fe], [Ba/Fe], and [Sr/Fe]. 
  In the previous sections, we have shown that in our sample of metal-poor stars with [Fe/H] close to --3.0\\ 
$\bullet$ there is a good correlation among abundances of the neutron-capture elements  belonging to the first peak (Fig. \ref{firstH} and Fig. \ref{firstFe});\\
$\bullet$ the scatter of [Sr/Ba] increases when [Ba/Fe] decreases (Fig. \ref{basr}{\tiny B}) whereas [Eu/Ba] remains constant (Fig. \ref{basr}{\tiny A}); and\\
$\bullet$ many metal-poor stars (8 out of 11) have an abundance pattern of heavy elements that is clearly different from the main r-process predictions as represented by the abundance pattern in CS\,31082-001  (Fig. \ref{pat2}, \ref{pat3}, and \ref{pat4}).

Since at this low metallicity a contamination by classical s-process yields from typical AGB stars was thought to be very improbable, the abundance pattern of HD\,122563 was first explained by a peculiar process producing all the elements of the first and second peak: weak r-process in \citet{HondaAI07}), truncated r-process in \citet{BoydFM12}, and also s-process in rotating massive VMP  stars \citep{FrischknechtHP16} (see also Limongi \& Chieffi (2015)\footnote{\tt \scriptsize https://ia2-owncloud.oats.inaf.it/index.php/s/U1yJ03hqz5T67va}).

\begin{figure}
\resizebox{8.4cm}{4.0cm}                   
{\includegraphics {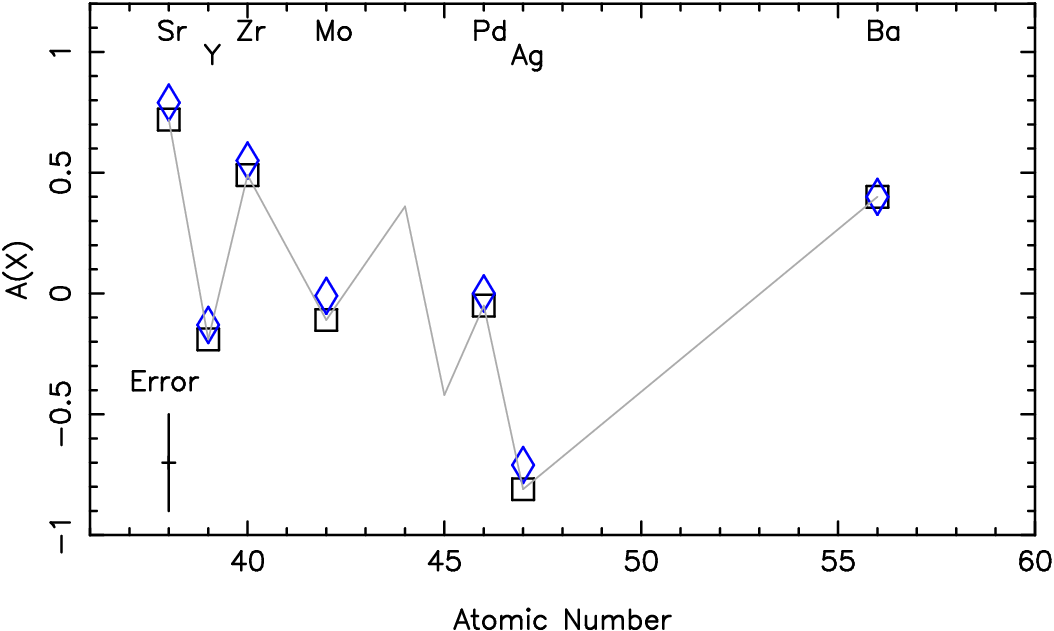}}
\caption[]{Simulation of the effect of a pollution on the r-rich star CS\,31082-001. We assume that a star such as CS31082-001 undergoes a pollution similar in atom numbers per $10^{12}$ hydrogen atoms as the pollution undergone by CS\,22897-008. The black squares represent the abundances before pollution and the diamonds indicate the abundances after pollution. In such a star, the effect of the pollution is almost not visible.}
\label {pollu}
\end{figure}

The fact that the spread in [Sr/Ba] increases when [Ba/Fe] decreases  (Fig. \ref{basr}{\tiny B}) rather suggests an enrichment in heavy elements of the matter in two independent steps. 
A first enrichment in pure r-process elements (as seen in CS\,3102-001 and in the stars in Fig. \ref{pat1}) and, independently, an enrichment of mainly the first peak elements \citep[as also suggested by ][]{CescuttiCH13,Roederer17}.
This contamination by a second mechanism becomes visible and even dominates in stars that are originally r-poor such as HD\,122563 and also in particular CS\,22897-008 (Fig. \ref{pat3}). If an r-rich star undergoes this same contamination, the observed distribution of the neutron-capture elements abundances (and in particular the ratio [Sr/Ba]\,) is not significantly affected. We  assume (Fig.\ref{pollu}) that CS\,31082-001 undergoes a pollution similar (in number of atoms per $10^{12}$ atoms of hydrogen) to the pollution needed to explain in this context, the pattern of CS\,22897-008. The star CS\,31082-001 is already so rich in n-capture elements that the change in the observed distribution of the neutron-capture abundances is not significant. As a consequence, this second mechanism could be very frequent in the history of the metal enrichment of the Galaxy. 
An r-poor star with  [Sr/Ba] close to the pure r-process value (like BD--18\,5550) would not have undergone a significant contamination by the  second mechanism, unlike HD\,122563 and the stars in Figs. \ref{pat3} and \ref{pat4}.

In Fig. \ref{basr}{\tiny A}, at a given value of [Ba/Fe], the [Sr/Ba] ratio depends on the importance of the contamination of the matter by the second mechanism. When a star is r rich, a pollution similar to the pollution observed in some r-poor stars is not perceptible and the ratio [Sr/Ba] remains constant. Thus at a given metallicity (here [Fe/H]=--3.0) the scatter of [Sr/Ba] increases when [Ba/Fe] decreases.

% Figure 10

% Figure 11
\begin{figure}
\resizebox{8.4cm}{5.0cm}                   
{\includegraphics {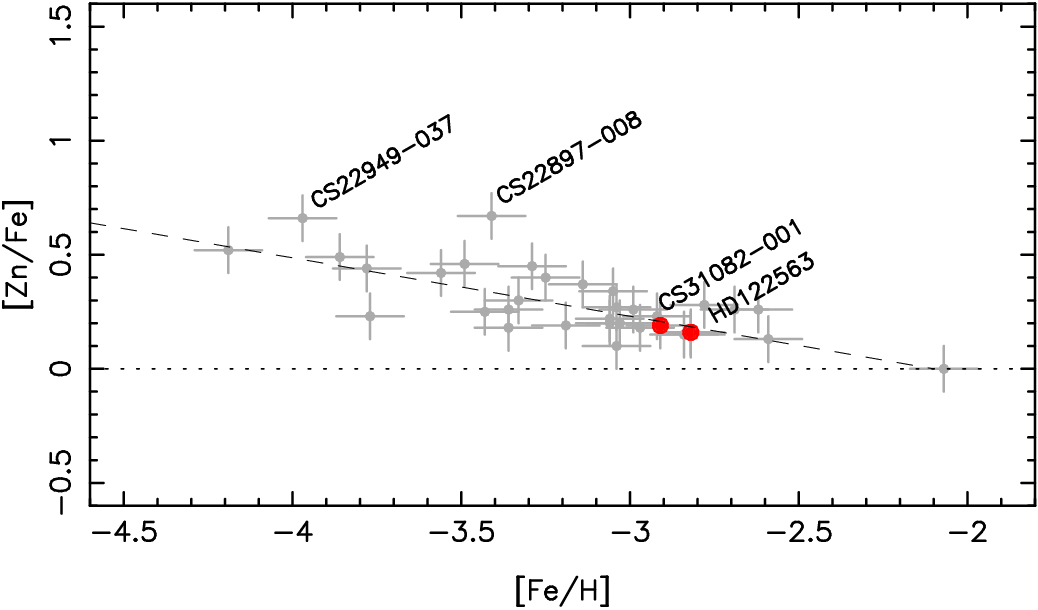}}
\caption[]{Zn abundance 
in the VMP and EMP giants here studied. The position of the two reference stars CS\,31082-001 (r-rich) and HD\,122563 (r-poor but Sr-rich) stars are indicated by red filled circles and both have a normal position for their metallicity. On the contrary, CS\,22897-008 and the CEMP star CS\,22949-037 seem to be Zn-rich.}
\label {Znab}
\end{figure}

% Figure 12
\begin{figure}
\begin{center}
\resizebox{6cm}{3.5cm}                  
{\includegraphics {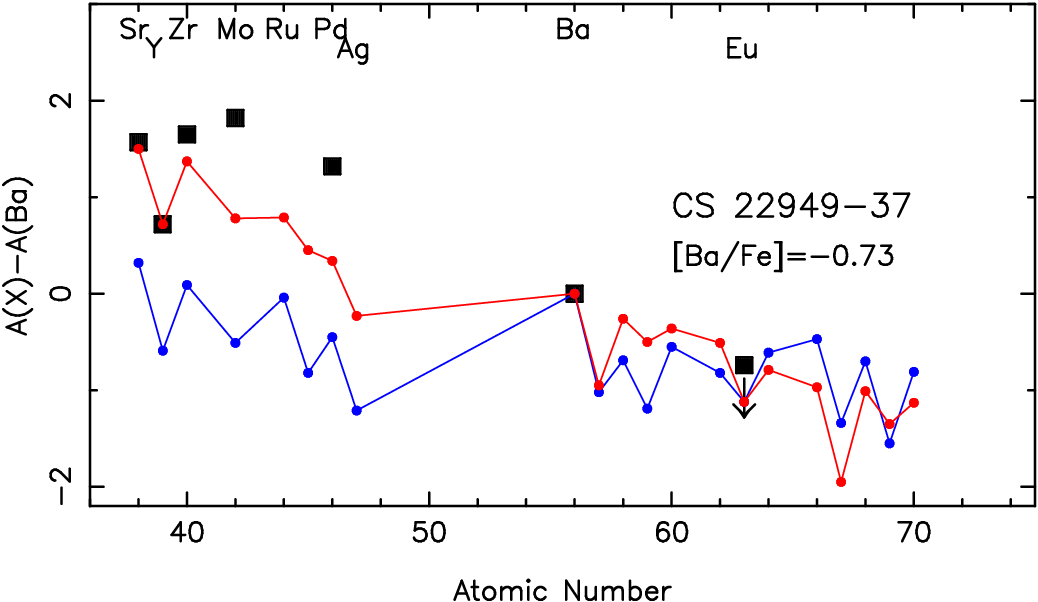}}
\caption[]{Abundance pattern of the neutron capture elements in the CEMP star CS\,22949-037 compared to CS\,31082-001 (blue line) and HD\,122563 (red line).}
\label{pat22949-37}
\end{center}
\end{figure}

\subsection{CEMP star CS\,22949-037}
In addition to the 11 classical VMP-EMP stars analysed here, we also considered the carbon-rich star CS\,22949-037 (Table \ref{abund1}, and \ref{abund2}), previously analysed in the context of the LP First Stars \citep{DepagneHS02}.  This star is EMP\ (with [Fe/H] =-3.97),  N-rich, and somewhat Zn-rich, but Ba-poor (Fig.\ref{Znab}). Since the abundance of Zr was not measured in \citet{DepagneHS02}, we measured the Zr lines in addition to Mo and Pd. In  Fig. \ref{firstFe}  this star stands out from the mean relations between the first peak elements observed in the EMP-VMP stars of our sample.  Although the Y abundance seems to be normal, obviously, in this star, the abundances of  Zr, Mo, and Pd are enhanced.
%quoting Norris et al. (2013), quoting himself the data of the LP-First stars  XXXX 2013 

The pattern of the element abundances (Fig. \ref{pat22949-37}) is somewhat similar to the pattern of HD\,122563 for Sr, Y, and Zr, but with higher abundances of  Mo and Pd. 
This star is included in the paper by \citet{FrischknechtHP16} (their section 6.2.2 and Table 8). These authors suggest (from only the few abundances available at this time, and mainly the [Sr/Ba] ratio) that the peculiar pattern of CS\,22949-037 could be explained by a FRMS model.
It would be interesting to check whether in particular, Zr, Mo, and Pd could be simultaneously explained. 

\section{Conclusions}

The general distribution of the heavy elements in space and time, shows a very complex picture. It seems that it is possible to explain our detailed observations by an enrichment in n-capture elements in two different events \citep[as proposed by][]{QianWas01,QianWas07,HansenMA14}.\\ 
In one event the natal cloud would be enriched by pure r-process  products of the primitive supernovae.\\
In another event this material would be enriched in (mainly) the first peak elements.
However, we stress that the data do not suggest in which order (if any) these two events occurred.\\
We then explain the nature of the second process.\\  

i) Following for example \citet{Wanajo13} various processes in supernovae (ECsupernovae) could overproduce the first peak elements (weak r-process), their production showing some variety. This variety is proposed by \citet{AokiIA17} to explain  the abundance patterns of the five galactic stars they observed (in the range $\rm -2.85 < [Fe/H] < -2.1$) that are similar to the pattern of the intermediate stars in our Fig.\ref{pat2}.\\

\indent ii) Also the production of first peak elements during the pulsations of AGB or SAGB \citep{BisterzoTG14}  stars could be invoked \citep{SiqueiraCB16}. 
This process could be important following \citet{BisterzoTW16}.\\

\indent iii) In a recent paper, \citet{FrischknechtHP16} provided new data about FRMS stars, some of these were able to produce a high proportion of first peak elements.
In this case, the large amount of neutrons, produced by the rapid rotation of the massive star, could transform the iron seed nuclei into heavy elements of mainly the first peak. The elements heavier than the first peak (and in particular Ba) would not be produced (or only in very small amounts) by such a rotational process.
This interesting theory has been recently confronted to the measurement of a star rich in first-peak neutron-capture elements (ROA 276) in the globular cluster $\rm \Omega ~Cen$. This peculiar star also presents an overabundance of Cu and Zn that is very well represented by the very specific pattern produced by the FRMS model. 
Unfortunately, in our VMP or EMP stars the abundance of Cu cannot be determined because the Cu lines in the visible, are too weak. But the abundance of Zn was measured in all the stars here studied \citep{CayrelDS04}. In HD\,122563 (Fig. \ref{Znab}), the ratio [Zn/Fe] is found to be normal, that is the same as the other stars of the same metallicity. But in this figure the more extreme star CS\,22897-008 seems to be Zn-rich for its metallicity.\\

\indent iv) One more possibility, following \citet{NishimuraST17}, would be the ``intermediate r-process in core-collapse supernovae driven by the magneto-rotational instability''. 
%\LEt{It is okay to keep the quotation marks if this is a quotation.}
This process also produces also an excess of Zn and could be also invoked to explain the peculiar pattern of CS\,22897-008.\\

\indent v) \citet{BanerjeeQH17} suggested that a significant number of protons could be ingested into the He shell at the end of the life of massive EMP stars (including zero-metal stars) of about 20 to 30$M_{\odot}$. This proton ingestion would induce a neutron flux that would react with primary C and O produced by He burning. Such nucleosynthesis would produce mainly Sr along with elements up to around Ba.

\begin {acknowledgements} 
%This work was partially supported by XXX XXX XXX,
BB acknowledges partial financial support from Fapesp, CNPq, and CAPES. 
\end {acknowledgements}

\bibliographystyle{aa}

\begin{thebibliography}{}

\bibitem[Alvarez \& Plez(1998)]{AlvarezP98}
Alvarez R., Plez B., 1998, A\&A 330, 1109

\bibitem[Aoki et al.(2017)]{AokiIA17} 
Aoki, M., Ishimaru, Y., Aoki, W., \& Wanajo, S.\ 2017, \apj, 837, 8 

\bibitem[Ballester et al.(2000)]{BallesterMB00} 
Ballester, P., Modigliani, A., Boitquin, O., et al.\ 2000, The Messenger, 101, 31 

\bibitem[Banerjee et al.(2017)]{BanerjeeQH17} 
Banerjee, P., Qian, Y.-Z., \& Heger, A.\ 2017, arXiv:1711.05964v1 

\bibitem[Bisterzo et al.(2014)]{BisterzoTG14} 
Bisterzo, S., Travaglio, C., Gallino, R., Wiescher, M., \& K{\"a}ppeler, F.\ 2014, \apj, 787, 10 

\bibitem[Bisterzo et al.(2016)]{BisterzoTW16} 
Bisterzo, S., Travaglio, C., Wiescher, M., et al.\ 2016, Journal of Physics Conference Series, 665, 012023 

\bibitem[Bisterzo et al.(2017)]{bisterzo17}
Bisterzo S, Travaglio, Wiescher M, K\"appeler F, Gallino R.
2017, \apj,  835, 97

\bibitem[Bonifacio et al.(2009)]{BonifacioSC09} 
Bonifacio, P., Spite, M., Cayrel, R., et al.\ 2009, \aap, 501, 519

\bibitem[Boyd et al.(2012)]{BoydFM12} 
Boyd, R.~N., Famiano, M.~A., Meyer, B.~S., et al.\ 2012, \apjl, 744, L14 
\bibitem[Cayrel et al.(2004)]{CayrelDS04} 
Cayrel, R., Depagne, E., Spite, M., et al.\ 2004, \aap, 416, 1117 

\bibitem[Cescutti et al.(2013)]{CescuttiCH13} 
Cescutti, G., Chiappini, C., Hirschi, R., Meynet, G., \& Frischknecht, U.\ 2013, \aap, 553, A51 

\bibitem[Cescutti \& Chiappini(2014)]{CescuttiChiap14} 
Cescutti, G., \& Chiappini, C.\ 2014, \aap, 565, A51 

%\bibitem[Choplin et al.(2017)]{ChoplinHM17} 
%Choplin, A., Hirschi, R., Meynet, G., \& Ekstr{\"o}m, S.\ 2017, \aap, 607, L3 

\bibitem[Choplin et al.(2017)]{ChoplinEM17} 
Choplin, A., Ekstr{\"o}m, S., Meynet, G., Maeder, A., Georgy, C., Hirschi, R.\ 2017, \aap, 605, A63 

\bibitem[Cowan et al.(2011)]{CowanRS11} 
Cowan, J.~J., Roederer, I.~U., Sneden, C., \& Lawler, J.~E.\ 2011, RR Lyrae Stars, Metal-Poor Stars, and the Galaxy, 5, 223 

\bibitem[Dekker et al.(2000)]{DekkerDK00} 
Dekker, H., D'Odorico, S., Kaufer, A., Delabre, B., \& Kotzlowski, H.\ 2000, \procspie, 4008, 534 

\bibitem[Depagne et al.(2002)]{DepagneHS02} 
Depagne, E., Hill, V., Spite, M., et al.\ 2002, \aap, 390, 187 

\bibitem[Fran{\c c}ois et al.(2007)]{FrancoisDH07} 
Fran{\c c}ois, P., Depagne, E., Hill, V., et al.\ 2007, \aap, 476, 935 

\bibitem[Freiburghaus et al.(1999)]{FreiburghausRT99} 
Freiburghaus, C., Rosswog, S., \& Thielemann, F.-K.\ 1999, \apjl, 525, L121 

\bibitem[Frischknecht et al.(2016)]{FrischknechtHP16} 
Frischknecht, U., Hirschi, R., Pignatari, M., et al.\ 2016, \mnras, 456, 1803 

\bibitem[Goriely et~al.(2015)]{goriely15}
Goriely S, Bauswein A, Janka H-T. 2015, \apj, 738, L32

\bibitem[Grevesse \& Sauval(2000)]{GrevesseSauval00} 
Grevesse, N., \& Sauval, A.~J.\ 2000, Origin of Elements in the Solar System, Implications of Post-1957 Observations, 261 

\bibitem[Gustafsson et al.(1975)]{GustafssonBE75}
Gustafsson B., Bell R. A., Eriksson K., Nordlund \AA., 1975, A\&A, 42, 407 

\bibitem[Gustafsson et al.(2003)]{GustafssonEE03}
Gustafsson B., Edvardsson B., Eriksson K., et al. 2003, in Stellar Atmosphere
Modeling, ed. I. Hubeny, D. Mihalas, \& K. Werner, ASP Conf. Ser., 288, 331 

\bibitem[Gustafsson et al.(2008)]{GustafssonEE08}
Gustafsson B., Edvardsson B., Eriksson K., Graae-J\o rgensen U., Nordlund\AA., Plez B., 2008, A\&A, 486, 951

\bibitem[Hansen \& Primas(2011)]{HansenP11} 
Hansen, C.~J., \& Primas, F.\ 2011, \aap, 525, L5 

\bibitem[Hansen et al.(2012)]{HansenPH12} 
Hansen, C.~J., Primas, F., Hartman, H., et al.\ 2012, \aap, 545, A31 

\bibitem[Hansen et al.(2014)]{HansenAC14} 
Hansen, C.~J., Andersen, A.~C., \& Christlieb, N.\ 2014, \aap, 568, A47 

\bibitem[Hansen et al.(2014)]{HansenMA14} 
Hansen, C.~J., Montes, F., \& Arcones, A.\ 2014, \apj, 797, 123 

\bibitem[Hill et al.(2002)]{HillPC02} 
Hill, V., Plez, B., Cayrel, R., et al.\ 2002, \aap, 387, 560 

\bibitem[Hill et al.(2016)]{2016HillCB}
Hill, V., Christlieb, N., Beers, T.~C., et al.\ 2016, arXiv:1608.07463

\bibitem[Hirschi et al.(2007)]{HirschiMM07} Hirschi, R., Maeder, A., Meynet, G., Chiappini, C., \& Ekstr{\"o}m, S.\ 2007, EAS Publications Series, 24, 263 

\bibitem[Honda et al.(2006)]{HondaAI06} 
Honda, S., Aoki, W., Ishimaru, Y., Wanajo, S., \& Ryan, S.~G.\ 2006, \apj, 643, 1180 

\bibitem[Honda et al.(2007)]{HondaAI07} 
Honda, S., Aoki, W., Ishimaru, Y., \& Wanajo, S.\ 2007, \apj, 666, 1189 

\bibitem[Jacobson et al.(2015)]{JacobsonKF15} 
Jacobson, H.~R., Keller, S., Frebel, A., et al.\ 2015, \apj, 807, 171

%\bibitem[Jones et al.(2016)]{JonesRH16} 
%Jones, S., Ritter, C., Herwig, F., et al.\ 2016, \mnras, 455, 3848 

\bibitem[K\"appeler et~al.(2011)]{kappeler11}
K\"appeler F, Gallino R, Busso M, Picchio G, 
Raiteri CM. 1990, Rev. Mod. Phys., 83, 157

\bibitem[Kasen et al.(2017)]{KasenMB17} 
Kasen, D., Metzger, B., Barnes, J., Quataert, E., \& Ramirez-Ruiz, E.\ 2017, \nat, 551, 80 

\bibitem[Lawler et al.(2004)]{LawlerSC04}
Lawler, J.E., Sneden, C., Cowan, J.J., et al.\ 2004, \apj, 604, 850

%\bibitem[Limongi et al.(2000)]{LimongiSC00} 
%Limongi, M., Straniero, O., \& Chieffi, A.\ 2000, \apjs, 129, 625 

\bibitem[Lodders et al.(2009)]{LoddersPG09}
Lodders, K., Palme H., \& Gail, H.P. 2009, In Landolt-Bornstein, New Series, Vol. VI/4B, Chap. 4.4, J.E. Truemper (ed.), Berlin, Heidelberg, New York: Springer-Verlag, p. 560 

\bibitem[Mashonkina et al.(2010)]{MashonkinaCB10} 
Mashonkina, L., Christlieb, N., Barklem, P.~S., et al.\ 2010, \aap, 516, A46 

\bibitem[Masseron et al.(2010)]{MasseronJP10} 
Masseron, T., Johnson, J.~A., Plez, B., et al.\ 2010, \aap, 509, A93 

\bibitem[Meynet et al.(2006)]{MeynetEM06} 
Meynet, G., Ekstr{\"o}m, S., \& Maeder, A.\ 2006, \aap, 447, 623 

\bibitem[Montes et al.(2007)]{MontesBC07} 
Montes, F., Beers, T.~C., Cowan, J., et al.\ 2007, \apj, 671, 1685 

\bibitem[Nishimura et al.(2017)]{NishimuraST17} 
Nishimura, N., Sawai, H., Takiwaki, T., Yamada, S., \& Thielemann, F.-K.\ 2017, \apjl, 836, L21 

\bibitem[Norris et al.(2002)]{NorrisRB02} 
Norris, J.~E., Ryan, S.~G., Beers, T.~C., Aoki, W., \& Ando, H.\ 2002, \apjl, 569, L107 

\bibitem[Peterson(2011)]{Peterson11} 
Peterson, R.~C.\ 2011, \apj, 742, 21 

\bibitem[Peterson(2013)]{Peterson13} 
Peterson, R.~C.\ 2013, \apjl, 768, L13 

\bibitem[Placco et al.(2017)]{PlaccoHF17} 
Placco, V.~M., Holmbeck, E.~M., Frebel, A., et al.\ 2017, arXiv:1706.02934 

\bibitem[Plez(2012)]{Plez12}
Plez B., 2012, ascl.soft05004P, http://adsabs.harvard.edu/abs/2012ascl.soft05004P

\bibitem[Roederer(2013)]{Roederer13} 
Roederer, I.~U.\ 2013, \aj, 145, 26 

\bibitem[Roederer(2017)]{Roederer17} 
Roederer, I.~U.\ 2017, \apj, 835, 23 

\bibitem[Qian \& Wasserburg(2001)]{QianWas01} 
Qian, Y.-Z., \& Wasserburg, G.~J.\ 2001, \apj, 549, 337 

\bibitem[Qian \& Wasserburg(2007)]{QianWas07} 
Qian, Y.-Z., \& Wasserburg, G.~J.\ 2007, \physrep, 442, 237

\bibitem[Siqueira Mello et al.(2012)]{SiqueiraBS12} 
Siqueira Mello, C., Barbuy, B., Spite, M., \& Spite, F.\ 2012, \aap, 548, A42 

\bibitem[Siqueira Mello et al.(2013)]{SiqueiraSB13} 
Siqueira Mello, C., Spite, M., Barbuy, B., et al.\ 2013, \aap, 550, A122 
\bibitem[Siqueira Mello et al.(2014)]{SiqueiraHB14} 
Siqueira Mello, C., Hill, V., Barbuy, B., et al.\ 2014, \aap, 565, A93 

\bibitem[Siqueira-Mello et al.(2015)]{SiqueiraAB15} 
Siqueira-Mello, C., Andrievsky, S.~M., Barbuy, B., et al.\ 2015, \aap, 584, A86 

\bibitem[Siqueira-Mello et al.(2016)]{SiqueiraCB16} 
Siqueira-Mello, C., Chiappini, C., Barbuy, B., et al.\ 2016, \aap, 593, A79 

\bibitem[Sneden et al.(2003)]{SnedenCL03} 
Sneden, C., Cowan, J.~J., Lawler, J.~E., et al.\ 2003, \apj, 591, 936 

\bibitem[Sneden et al.(2008)]{SnedenCG08} 
Sneden, C., Cowan, J.~J., \& Gallino, R.\ 2008, \araa, 46, 241 

\bibitem[Spite et al.(2014)]{SpiteSB14} 
Spite, M., Spite, F., Bonifacio, P., et al.\ 2014, \aap, 571, A40 

\bibitem[Spite et al.(2005)]{SpiteCP05} 
Spite, M., Cayrel, R., Plez, B., et al.\ 2005, \aap, 430, 655 

\bibitem[Spite \& Spite(2014)]{SpiteS14} 
Spite, M., \& Spite, F.\ 2014, Astronomische Nachrichten, 335, 65 

\bibitem[Starkenburg et al.(2014)]{StarkenburgSM14} 
Starkenburg, E., Shetrone, M.~D., McConnachie, A.~W., \& Venn, K.~A.\ 2014, \mnras, 441, 1217 

\bibitem[Vangioni et al.(2016)]{VangioniGD16} 
Vangioni, E., Goriely, S., Daigne, F., Fran{\c c}ois, P., \& Belczynski, K.\ 2016, \mnras, 455, 17 

\bibitem[Wanajo et al.(2011)]{WanajoJM11} 
Wanajo, S., Janka, H.-T., \& M{\"u}ller, B.\ 2011, \apjl, 726, L15 

\bibitem[Wanajo(2013)]{Wanajo13} 
Wanajo, S.\ 2013, \apjl, 770, L22 

\bibitem[Wanajo et~al.(2014)]{wanajo14} 
Wanajo S, Sekiguchi Y, Nishimura N, %Kiuchi K, Kyutoku K, Shibata M. 
et~al. 2014, \apj, 789, L39 

\bibitem[Winteler et al.(2012)]{WintelerKP2012} 
Winteler, C., K{\"a}ppeli, R., Perego, A., et al.\ 2012, \apjl, 750, L22 
\bibitem[Yong et al.(2017)]{YongND17} 
Yong, D., Norris, J.~E., Da Costa, G.~S., et al.\ 2017, \apj, 837, 176 



\end{thebibliography}
{}

\end{document}